\documentclass[11pt]{article}
\usepackage{amssymb,latexsym,amsmath, amsthm}
\usepackage{amsfonts, graphics,color}
\usepackage{mathtools}
\usepackage{hyperref} 
\usepackage{breakurl} 
\usepackage[american]{babel}
\usepackage{csquotes}
\usepackage{enumitem}
\usepackage{dsfont}
\usepackage[utf8]{inputenc}

\usepackage[top=1in, bottom=1.25in, left=1.25in, right=1.25in]{geometry}

\usepackage{subcaption} 
\usepackage{float}

\def\R{\mathbb R}

\def\N{\mathbb N}

\def\T{{\mathcal T}}

\def\Rmax{R_{\mathrm{max}}}
\def\Rmin{R_{\mathrm{min}}}
\newcommand\Lmax{L_{\mathrm{max}}}

\def\Ri{{\mathcal R }}

\def\L{{\mathcal L}}

\newcommand\DEL{\mathbb{D}}
\newcommand{\Tmin}{T_{\mathrm{min}}}
\newcommand{\Tmax}{T_{\mathrm{max}}} 
\newcommand{\sigmaess}{\sigma_{\mathrm{ess}}}

\newcommand\Emin[1]{E^{\mathrm{min}}_{#1}}

\newcommand{\Gref}{G^{\mathrm{ref}}}

\let\originalleft\left
\let\originalright\right
\renewcommand{\left}{\mathopen{}\mathclose\bgroup\originalleft}
\renewcommand{\right}{\aftergroup\egroup\originalright}

\newcommand*\diff{\mathop{}\!\mathrm{d}} 

\newcommand{\Etot}{E_{\mathrm{tot}}}
\newcommand{\Ekin}{E_{\mathrm{kin}}}

\newcommand{\Elin}{E^{\mathrm{lin}}}
\newcommand{\Elinkin}{E^{\mathrm{lin}}_{\mathrm{kin}}}
\newcommand{\Elinpot}{E^{\mathrm{lin}}_{\mathrm{pot}}}
\newcommand{\Efree}{E^{\mathrm{free}}}

\renewcommand{\epsilon}{\varepsilon}

\usepackage{xcolor}
\usepackage[normalem]{ulem} 
\def\bcr{\begin{color}{red}}
\def\bcb{\begin{color}{blue}}
\def\bcc{\begin{color}{violet}}
\definecolor{darkgreen}{RGB}{0,150,0}
\def\bcg{\begin{color}{darkgreen}}
\def\ec{\end{color}}
\def\be{\begin{equation}}
\def\ee{\end{equation}}

\hypersetup{pdftitle={Straub-VPNumerics}}

\newtheorem{theorem}{Theorem}[section]

\newtheorem{observation}[theorem]{Observation}

\numberwithin{equation}{section}
\numberwithin{figure}{section}

\newcommand{\CC}{C\nolinebreak\hspace{-.05em}\raisebox{.4ex}{\tiny\bf +}\nolinebreak\hspace{-.10em}\raisebox{.4ex}{\tiny\bf +}}

\title{\vspace{-1.8em}Numerical experiments on stationary, oscillating, and damped spherical galaxy models}

\author{Christopher~Straub\vspace{0.15em}\\ 
  Department of Mathematics, University of Bayreuth, Germany\vspace{0.15em}\\
  \href{mailto:christopher.straub@uni-bayreuth.de}{christopher.straub@uni-bayreuth.de}}

\begin{document}
\maketitle
\vspace{-1.9em}

\begin{abstract}
	We numerically analyse solutions of the spherically symmetric gravitational Vlasov-Poisson system close to compactly supported stable steady states.
	We observe either partially undamped oscillations or macroscopically damped solutions.
	We investigate for many steady states to which of these behaviours they correspond.
	A linear relation between the exponents of polytropic steady states and the qualitative behaviour close to them is identified.
	Undamped oscillations are also observed around not too concentrated King models and around all shells with a sufficiently large vacuum region.
	We analyse all solutions both at the non-linear and linearised level and find that the qualitative behaviours are identical at both.
	To relate the observed phenomena to theoretical results, we further include a comprehensive numerical study of the radial particle periods in the equilibria.
\end{abstract}

\vspace{-1.5em}

\tableofcontents

\section{Introduction}\label{sc:intro}

A classical model for the evolution of star clusters or galaxies is the three-dimensional gravitational Vlasov-Poisson system
\begin{equation}\label{eq:vlasov}
	\partial_tf+v\cdot\partial_xf-\partial_xU\cdot\partial_vf=0,
\end{equation}
\begin{equation}\label{eq:poisson}
	\Delta U = 4\pi\rho,\qquad\lim_{|x|\to\infty}U(t,x)=0,
\end{equation}
\begin{equation}\label{eq:rho}
	\rho(t,x)=\int_{\R^3}f(t,x,v)\diff v.
\end{equation}
The main unknown is the phase space density function $f=f(t,x,v)\geq0$ of the configuration depending on the time $t\in\R$, spatial position $x\in\R^3$, and velocity $v\in\R^3$.
Its evolution is determined by the Vlasov equation~\eqref{eq:vlasov}; dots~$\cdot$ denote the Euclidean scalar product.
The gravitational potential $U=U(t,x)$ of the configuration is determined by the Poisson equation~\eqref{eq:poisson}; the boundary condition included into~\eqref{eq:poisson} corresponds to an isolated system.
These equations are coupled by imposing that the mass density $\rho=\rho(t,x)$ is given by~\eqref{eq:rho}.
For background on this system we refer to~\cite{BiTr,Gl96,Re07}.

We consider this system in spherical symmetry, i.e., we assume
\begin{equation}\label{eq:defsphsymm}
	f(t,Ax,Av)=f(t,x,v),\qquad t\in\R,\;x,v\in\R^3,\; A\in\mathrm{SO}(3).
\end{equation}
This allows us to write $f(t)$ as a function of the reduced variables
\begin{equation}\label{eq:defrwL}
	r=|x|,\quad w=\frac{x\cdot v}r,\quad L=|x\times v|^2,
\end{equation}
which are the spatial radius~$r$, the radial velocity~$w$, and the squared modulus of the angular momentum~$L$.
In addition, we can write $\rho=\rho(t,r)$ and $U=U(t,r)$.
The Vlasov-Poisson system in spherical symmetry describes the evolution of globular clusters or spheroidal galaxies.

This system possesses a plethora of steady states, i.e., stationary solutions, describing star clusters in an equilibrium state.
By Jeans' theorem~\cite[Thm.~2.2]{BaFaHo86}, any steady state~$f_0=f_0(x,v)=f_0(r,w,L)$ is a function of the particle energy
\begin{equation}\label{eq:defE}
	E(x,v)\coloneqq\frac12|v|^2+U_0(x)
\end{equation}
and the (squared) modulus of the angular momentum $L=L(x,v)$ defined in~\eqref{eq:defrwL}.
More precisely, $f_0$ is of the form 
\begin{equation}\label{eq:f0varphi}
	f_0(x,v)=\varphi(E(x,v),L(x,v)),
\end{equation}
where $\varphi\colon\R^2\to[0,\infty[$ is the microscopic equation of state of the steady state and~$U_0$ is its gravitational potential.
It is well-understood which choices of~$\varphi$ correspond to steady states with finite mass and compact support, cf.~\cite{RaRe2013} and the references therein.
In Section~\ref{sc:stst}, we present the microscopic equations of state~$\varphi$ used here, which correspond to some of the most commonly used steady states in the literature: isotropic and anisotropic polytropes, King models, and shells. 
A broad overview of stationary solutions of the Vlasov-Poisson system in spherical symmetry, including many more equilibria than covered here, can be found in~\cite[Sc.~4.3]{BiTr}.

Because star clusters are frequently exposed to small perturbations, e.g., gravitational forces of other nearby stellar objects, only stable steady states can persist in nature.
It was first argued by Antonov~\cite{An62} that (certain) steady states satisfying
\begin{equation}\label{eq:varphiprimeneg}
	\partial_E\varphi<0
\end{equation}
on their support are stable at a linearised level.
The condition~\eqref{eq:varphiprimeneg} became known as {\em Antonov's stability criterion}.
It is a natural condition from a physics point of view~\cite{ZeNo71}; it means that the concentration of ever more energetic particles is decreasing within the equilibrium.
Further linear stability results were proven in~\cite{BaMoRe95,DoFeBa73,KaSy85}.
The actual non-linear stability of all (suitable) steady states satisfying~\eqref{eq:varphiprimeneg} was later shown in~\cite{GuRe99,LeMeRa12}; see~\cite{Mo13} or~\cite[Sc.~4]{Re23} for a review.\footnote{It should again be emphasised that we are considering the full spherically symmetric setting. In general, spherically symmetric steady states of the form~\eqref{eq:f0varphi} satisfying~\eqref{eq:varphiprimeneg} are not stable against non spherically symmetric perturbations, cf.~\cite[Sc.~6]{Me99}.}

In order to understand the behaviour of star clusters in reality, solutions of the Vlasov-Poisson system close to stable steady states must be considered.
In the astrophysics literature, numerical analyses of such solutions commenced decades ago~\cite{BaGoHu86,He72,LeCoBi93,LoGe88,Me87,MiSm94,Na00,PeAlAlSc96,RaRe2018,SePr98,Sw93,WaMu97,WaRyMu93,We94}.
It was always observed that 
either parts of the solutions oscillate around the equilibrium  or that the solutions converge to the equilibrium on a macroscopic level.
First theoretical explanations for such damping mechanisms were discussed in~\cite{LyBe62,LyBe67}, explanations and examples (in settings different to the one considered here) for oscillatory solutions in~\cite{LoGe88,Ma90,Va03}.
More recently, mathematical investigations regarding these phenomena have been conducted in~\cite{HaReScSt23,HaReSt21,Ku21,MoRiBo23,RiSa2020}.
Nevertheless, the existence of undamped oscillations or damping cannot yet be rigorously proven for any steady state of the above form.
The aim of the present work is to contribute to the closing of this gap.
In order to achieve this, various quantities/behaviours are investigated here, the understanding of which appears to be particularly relevant for the advancement of theoretical results.

An important quantity appearing in various theoretical studies~\cite{HaReScSt23,HaReSt21,Ku21,Ma90,MoRiBo23,RiSa2020} is the radial period function of a steady state.
This function describes the radial periods of the particles within the equilibrium and its properties are intimately related to the dynamics around the steady state, cf.\ Section~\ref{ssc:Ttheory}.
Some properties of this function -- like its regularity and boundedness -- can be established rigorously, cf.~\cite[App.~B]{HaReSt21}, \cite[Ch.~3]{Ku21}, and~\cite[App.~B]{RiSa2020}.
We will review these results in Section~\ref{ssc:Ttheory}.
However, many other properties -- like characterising the particles with the longest radial periods within the equilibrium -- remain elusive.
In Section~\ref{ssc:Tnum} we numerically analyse in detail the radial period functions associated to the steady states introduced in Section~\ref{sc:stst}.
To our knowledge, this is the first such numerical study.
For many steady states we find that the longest radial periods correspond to the particles with the largest orbit, cf.\ Observation~\ref{obs:Tmaxcorner}.
In particular, we numerically verify the assumptions made in~\cite[Thm.~8.15 and Rem.~8.16]{HaReSt21} for a rigorous proof of the presence of oscillatory modes around some steady states.
In addition, for all steady states we find that the largest periods are attained by particles on the boundary of the steady state support, cf.\ Observation~\ref{obs:Tmaxboundary}.

In Section~\ref{sc:dyn} we then analyse the dynamics close to stable steady states through numerical simulations of the Vlasov-Poisson system after linearising it around the respective steady state.
The reason we focus on this linearised system is because all of the aforementioned theoretical works regarding the behaviour of solutions close to stable steady states exclusively concern the linearised setting.
Nonetheless, there have been only a few numerical studies of the linearised system; numerical analyses usually directly focus on the actual non-linear system.
Exceptions of this are~\cite{LeCoBi93,Me87,WaRyMu93}, where stationary solutions different to those considered here were analysed.
To close this gap, we mainly investigate the linearised system here.
We first introduce this system in Section~\ref{ssc:LVPtheory}.
The numerical simulations performed in Section~\ref{ssc:LVPnum} again confirm that solutions of the linearised Vlasov-Poisson system, in the case of a stable steady state, either oscillate partially undamped or are macroscopically damped.
We analyse in detail which steady states correspond to which of these behaviours, cf.\ Observations~\ref{obs:LVP_isopoly_oscvdamp}, \ref{obs:LVP_King}, \ref{obs:LVP_anisopoly}, and~\ref{obs:LVP_polyshell}.
We also review previous numerical investigations and observe that they are consistent with the observations made here.
Furthermore, we discuss the relations between the observed oscillation periods and the spectral considerations from the mathematical literature, cf.\ Observation~\ref{obs:LVP_isopoly_noembedded}.

In Section~\ref{sssc:LVPproofs} we discuss some approaches to rigorously show the presence of oscillatory modes.
In particular, we check the validity of a criterion derived in~\cite[Cor.~2.2]{Ku21} for the existence of an oscillatory mode around a stable steady state.
For this criterion, the longest radial particle period must be compared with macroscopic quantities of the steady state.
For certain steady states, we numerically show that the criterion is indeed satisfied.

We then numerically analyse solutions of the (non-linearised) Vlasov-Poisson system close to the same steady states as before in Section~\ref{ssc:VPnum}.
Our main goal is to check whether these solutions are accurately described by solutions of the respective linearised Vlasov-Poisson system.
We find that indeed they are.
More precisely, for any steady state for which we observed undamped oscillations at the level of the linearised Vlasov-Poisson system, we find that the solutions of the non-linear system close to the steady state oscillate undamped with the same period. 
If all solutions of the linearised Vlasov-Poisson system were macroscopically damped, the damping is also present for solutions of the non-linear system close to the respective steady state.
In this case, we even find that the damping rates at the linearised and non-linear level are identical. 
This is a remarkable property of the Vlasov-Poisson system because for other PDEs damping effects occur when transitioning from the linearised to the non-linear level~\cite{SoWe99}.

In Section~\ref{ssc:VPnum}, we also consider solutions of the respective pure transport equations. 
This equation arises when neglecting the influence of the gravitational response of the perturbation in the linearised Vlasov-Poisson system and it has often been studied as a first step towards damping results~\cite{ChLu22,LyBe62,MoRiBo22,RiSa2020}.
However, we find that the solutions of the pure transport equation differ qualitatively from those of the linearised and non-linear Vlasov-Poisson system, cf.\ Observation~\ref{obs:VP_lin}.

In Section~\ref{sc:out} we discuss several open research questions.
On the one hand, these contain theoretical problems which seem promising after the numerical analysis conducted here. 
On the other hand, we discuss numerical issues which ought to provide further insights into the questions addressed here.

The numerical methods used throughout this work and their accuracies are discussed in Appendix~\ref{sc:num}.
Computing a steady state corresponds to solving a scalar integro-differential equation.
The radial particle periods are determined by solving the characteristic system using the Runge-Kutta method.
Solutions of the non-linear Vlasov-Poisson system are evolved using a particle-in-cell scheme. 
The same numerical method has been used in~\cite{RaRe2018} to simulate the Vlasov-Poisson system and in~\cite{Praktikum20,GueStRe21} in the relativistic case.
Compared to $N$-body simulations, this method is more adapted to the Vlasov-Poisson system; see the convergence result~\cite{Sc87}.
In order to simulate solutions of the linearised Vlasov-Poisson system, we also use a suitable adaption of the particle-in-cell method.
We will, however, only outline the numerical methods here.
We instead prefer to make our code publicly available so that it can be checked, used, adapted, or extended by anyone.

\phantom{.}

\noindent
{\bf Acknowledgments.}
The majority of the findings presented here and the code used to obtain them originate in the author's doctoral thesis~\cite{St24}.

\section{Steady states}\label{sc:stst}

We consider steady states~$f_0$ given by~\eqref{eq:f0varphi} with microscopic equations of state~$\varphi$ of the form
\begin{equation}\label{eq:varphigeneral}
	\varphi(E,L)=\Phi(E_0-E)\,(L-L_0)_+^\ell,\qquad E,L\in\R,
\end{equation}
where the index~\enquote{$+$} denotes taking the positive part.
The energy dependency of the steady state is described by the prescribed function $\Phi\colon\R\to[0,\infty[$ and the parameter $E_0<0$.
We always consider energy dependency functions with $\Phi(\eta)=0$ for $\eta\leq0$ so that~$E_0$ plays the role of a cut-off energy.
In the case $\ell=0=L_0$, the steady state~$f_0$ depends only the particle energy~$E$ given by~\eqref{eq:defE} and is called {\em isotropic}~\cite[Sc.~4.2.1(a)]{BiTr}; we use the convention $b_+^0=1$ for $b\in\R$ in~\eqref{eq:varphigeneral} in this case.
We consider two classes of such steady states here. 
Firstly, the {\em isotropic polytropes}
\begin{equation}\label{eq:f0isopoly}
	f_0(x,v)=(E_0-E(x,v))_+^k,
\end{equation}
which correspond to energy dependency functions~$\Phi(\eta)=\eta_+^k$ with $0<k<\frac72$, cf.~\cite[Sc.~4.3.3(a)]{BiTr} for background on these models.
Secondly, the {\em King models}
\begin{equation}\label{eq:f0King}
	f_0(x,v)=(e^{E_0-E(x,v)}-1)_+,
\end{equation}
which correspond to $\Phi(\eta)=(e^\eta-1)_+$ and have first been proposed in~\cite{Ki66,Mi63,MiBo63} as adequate models for star clusters in an equilibrium state; see~\cite{Ki81} and~\cite[Sc.~4.3.3(c)]{BiTr} for further background on these models.

Steady states that not only depend on the particle energy~$E$ are called {\em anisotropic}.
A commonly studied class~\cite{BaFaHo86,Gu99-1,He72,Wo99} of such steady states are the {\em (anisotropic) polytropes}
\begin{equation}\label{eq:f0poly}
	f_0(x,v)=(E_0-E(x,v))_+^k\,L(x,v)^\ell,
\end{equation}
where $\ell>0$ and $0<k<3\ell+\frac72$ are prescribed polytropic exponents; with $\ell=0$ we recover the isotropic polytropes~\eqref{eq:f0isopoly}.
An extension of the polytropes can be obtained by choosing $L_0>0$ in~\eqref{eq:varphigeneral}.
The resulting steady states are of the form
\begin{equation}\label{eq:f0polyshell}
	f_0(x,v)=(E_0-E(x,v))_+^k\,(L(x,v)-L_0)_+^\ell
\end{equation}
and are referred to as {\em (polytropic) shells} with polytropic exponents $\ell>-\frac12$ and $0<k<3\ell+\frac72$; such steady states have been analysed in~\cite{HaReSt21,Re99-2,Sc06} among others.
The energy dependency function~$\Phi$ corresponding to~\eqref{eq:f0poly} and~\eqref{eq:f0polyshell} is the same as for~\eqref{eq:f0isopoly}.

\begin{figure}[h!]	
	\begin{center}
		\centering
		\includegraphics[width=\textwidth]{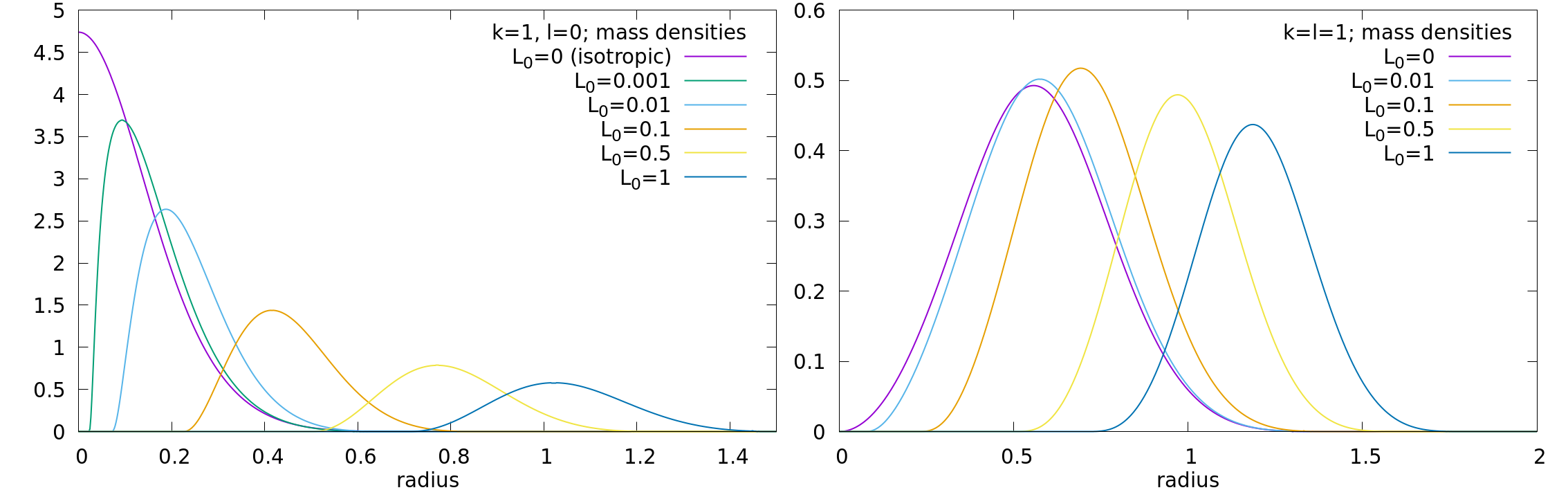}
	\end{center}
	\vspace*{-1.5em}
	\caption{Mass densities~$\rho_0$ of some polytropic shells~\eqref{eq:f0polyshell}, an anisotropic polytrope~\eqref{eq:f0poly}, and an isotropic polytrope~\eqref{eq:f0isopoly} with $\kappa=1$. 
		In the left panel we have chosen the polytropic exponents $(k,\ell)=(1,0)$ and $L_0\in\{0,\frac1{1000},\frac1{100},\frac1{10},\frac12,1\}$, the right panel corresponds to the polytropic exponents $(k,\ell)=(1,1)$ and $L_0\in\{0,\frac1{100},\frac1{10},\frac12,1\}$.}
	\label{fig:polyshell_radials}
\end{figure}

Because~$E$ and~$L$ are constant along the characteristic flow
\begin{equation}\label{eq:charsysxv}
	\dot x=v,\qquad\dot v=\partial_xU_0(x),
\end{equation}
in spherical symmetry, any function of the general form~\eqref{eq:f0varphi} automatically solves the Vlasov equation~\eqref{eq:vlasov}.
Hence, obtaining a steady state of the above form is reduced to solving the Poisson equation~\eqref{eq:poisson}.
By inserting~\eqref{eq:f0varphi} and~\eqref{eq:varphigeneral} into~\eqref{eq:rho} and expressing the Poisson equation in its radial form, a straight-forward calculation~\cite[p.~905]{RaRe2013} leads to the following equation for~$U_0$:
\begin{equation}\label{eq:U0poisson}
	U_0'(r)=\frac{4\pi}{r^2}\int_0^rs^{2\ell+2}g\left(E_0-U_0(s)-\frac{L_0}{2s^2}\right)\diff s,\qquad r>0,
\end{equation}
together with the boundary condition
\begin{equation}\label{eq:U0boundary}
	\lim_{r\to\infty}U_0(r)=0.
\end{equation}
The function~$g$ appearing in~\eqref{eq:U0poisson} describes the relation between~$\rho_0$ and~$U_0$ via
\begin{equation}\label{eq:rho0U0}
	\rho_0(r)=r^{2\ell}g\left(E_0-U_0(r)-\frac{L_0}{2r^2}\right),\qquad r>0,
\end{equation}
and is given by
\begin{equation}\label{eq:defg}
	g\colon\R\to\R,\;g(z)\coloneqq\begin{cases}c_\ell\int_0^z\Phi(\eta)\,\left(z-\eta\right)^{\ell+\frac12}\diff\eta,&\text{ if } z>0,\\0,&\text{ if } z\leq0,
	\end{cases}
\end{equation}
where $c_\ell\coloneqq2^{\ell+\frac32}\,\pi^{\frac32}\,{\Gamma(\ell+1)}\,{\Gamma(\ell+\frac32)^{-1}}$.
In the polytropic case $\Phi(\eta)=\eta_+^k$, it is possible to computed~$g$ explicitly, cf.~\cite[Ex.~4.1]{BaFaHo86}. 
To obtain a solution of \eqref{eq:U0poisson}--\eqref{eq:U0boundary} we proceed as in~\cite{BaFaHo86,RaRe2013} and consider the quantity $y\coloneqq E_0-U_0$ instead of~$U_0$, which makes the cut-off energy~$E_0$ part of the unknowns.
The equation for~$y$ reads
\begin{equation}\label{eq:ypoisson}
	y'(r)=-\frac{4\pi}{r^2}\int_0^rs^{2\ell+2}g\left(y(s)-\frac{L_0}{2s^2}\right)\diff s,\qquad r>0.
\end{equation}
We equip this integro-differential equation with the boundary condition
\begin{equation}\label{eq:yboundary}
	y(0)=\kappa
\end{equation}
for prescribed $\kappa>0$.
For any $\kappa>0$, there exists a unique solution $y\in C^2([0,\infty[)$ of~\eqref{eq:ypoisson}--\eqref{eq:yboundary} with $y'(0)=0$, cf.~\cite[p.~906]{RaRe2013}.
Moreover, for the choices of steady states presented above, $y_\infty\coloneqq\lim_{r\to\infty}y(r)\in]-\infty,0[$.
For the King models~\eqref{eq:f0King}, this is due to the compact-support lemma from~\cite[Lemma~3.1]{RaRe2013}.
In the polytropic case with $L_0=0$, i.e., for steady states of the form~\eqref{eq:f0isopoly} and~\eqref{eq:f0poly}, it is proven in~\cite[Lemma~5.3]{BaFaHo86} and~\cite{Sa40} that~$y$ possesses a zero which obviously leads to $y_\infty<0$.
This statement can then be extended to the case $L_0>0$ by a perturbation argument similar to~\cite[Thm.~1]{Re99-2}.
By defining $E_0\coloneqq y_\infty<0$ and $U_0\coloneqq E_0-y$, one obtains a solution of \eqref{eq:U0poisson}--\eqref{eq:U0boundary}.
Inserting~$U_0$ into~\eqref{eq:rho0U0} and~\eqref{eq:f0varphi} gives the mass density~$\rho_0$ and phase space density~$f_0$ of the resulting steady state, respectively. 
The negative cut-off energy~$E_0<0$ causes this steady state to have finite mass and compact support, i.e.,
\begin{equation}\label{eq:defM0}
	M_0\coloneqq\int_{\R^3\times\R^3}f_0(x,v)\diff(x,v)=\int_{\R^3}\rho_0(x)\diff x\in{]0,\infty[}
\end{equation} 
and 
\begin{equation}
	\Rmax\coloneqq\sup\{r\geq0\mid\rho_0(r)>0\}\in{]0,\infty[}.
\end{equation}
Furthermore, the arguments in~\cite[p.~163f.]{BaFaHo86} show that the radial support of the steady state is of the form
\begin{equation}
	\{r>0\mid\rho_0(r)>0\}={]\Rmin,\Rmax[}.
\end{equation}
In the case $L_0>0$, the inner radius~$\Rmin$ is positive and determined by~$L_0$ and~$\kappa$ via
\begin{equation}\label{eq:Rminkappa}
	\Rmin=\sqrt{\frac{L_0}{2\kappa}}\in{]0,\Rmax[}.
\end{equation}
This inner vacuum region is the reason why we refer to the steady states~\eqref{eq:f0polyshell} as shells.
In the case $L_0=0$, there holds $\Rmin=0$, i.e., the spatial steady state support is a ball.
In the anisotropic case~\eqref{eq:f0poly}, the mass density vanishes at the centre of the steady state by~\eqref{eq:rho0U0}, i.e., $\rho_0(0)=0$.
Notice that we only allow for positive values of the exponent~$\ell$ in the anisotropic polytropic case~\eqref{eq:f0poly}; negative~$\ell$ would result in the mass density~$\rho_0$ to become singular at the centre $r=0$.
In the isotropic case, the central density is positive and in one-to-one correspondence to the parameter~$\kappa$ via
\begin{equation}\label{eq:rho0centrekappa}
	\rho_0(0)=g(\kappa)>0;
\end{equation}
it follows by~\cite[p.~905]{RaRe2013} that~$g$ is strictly increasing on~$]0,\infty[$.
This also shows that, in the isotropic case, $[0,\Rmax]\ni r\mapsto\rho_0(r)$ is strictly decreasing.
In order to visualise the properties of the mass densities for the different classes of steady states, some examples for~$\rho_0$ are depicted in Figure~\ref{fig:polyshell_radials}.

\begin{figure}[h!]	
	\begin{center}
		\centering
		\includegraphics[width=\textwidth]{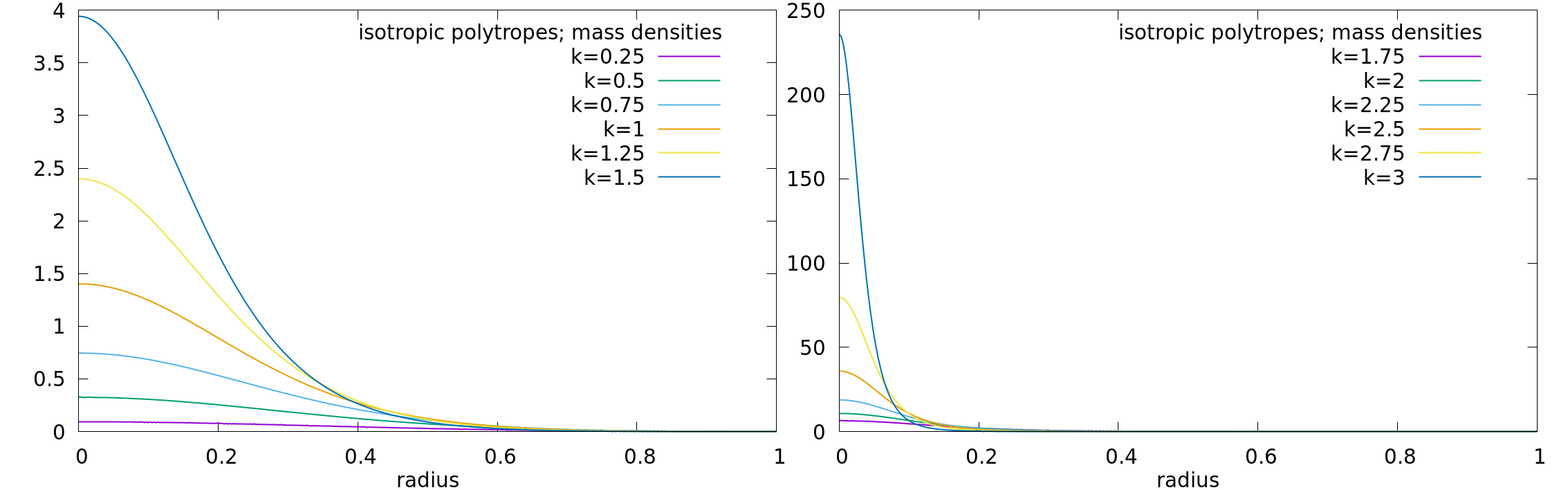}
	\end{center}
	\vspace*{-1.5em}
	\caption{Mass densities~$\rho_0$ of isotropic polytropes~\eqref{eq:f0isopoly} with $\Rmax=1$ and polytropic exponents $k\in\{\frac14,\frac12,\frac34,1,\frac54,\frac32\}$ (left panel) and $k\in\{\frac74,2,\frac94,\frac52,\frac{11}4,3\}$ (right panel).}
	\label{fig:isopoly_radials}
\end{figure}

In summary, the above shows that any $\kappa>0$ together with a prescribed ansatz -- determined by~$\Phi$, $\ell$, and~$L_0$ -- lead to a unique steady state.
Let us briefly discuss the role of the parameter $\kappa>0$.
In the polytropic cases~\eqref{eq:f0isopoly} and~\eqref{eq:f0poly}, varying~$\kappa>0$ corresponds to rescaling the steady state.
This scaling law originates from the study of equilibria of the Euler-Poisson system, where it is known as the Eddington-Ritter relation~\cite[p.~235]{Ro43}.
In the context of the Vlasov-Poisson system it was derived in~\cite[Sc.~3.3]{HaReSt21} and, in the isotropic case, in~\cite[Ch.~6]{Ku21}.
It follows by~\cite{DoRe01,HaReSt21} that such rescaling does not qualitatively change the behaviour of solutions close to the steady state; see also~\cite[Sc.~4]{RaRe2018} for a numerical study.
In these polytropic cases, we hence always choose the (unique) value for~$\kappa>0$ leading to $\Rmax=1$.
Some of the resulting isotropic polytropes are visualised in Figure~\ref{fig:isopoly_radials}.
The figure shows that the mass becomes rather concentrated around the centre for larger polytropic exponents~$k$. 
As such configurations are difficult to accurately handle numerically, we will henceforth restrict ourselves to isotropic polytropes with $0<k\leq3$.

\begin{figure}[h!]	
	\begin{center}
		\centering
		\includegraphics[width=\textwidth]{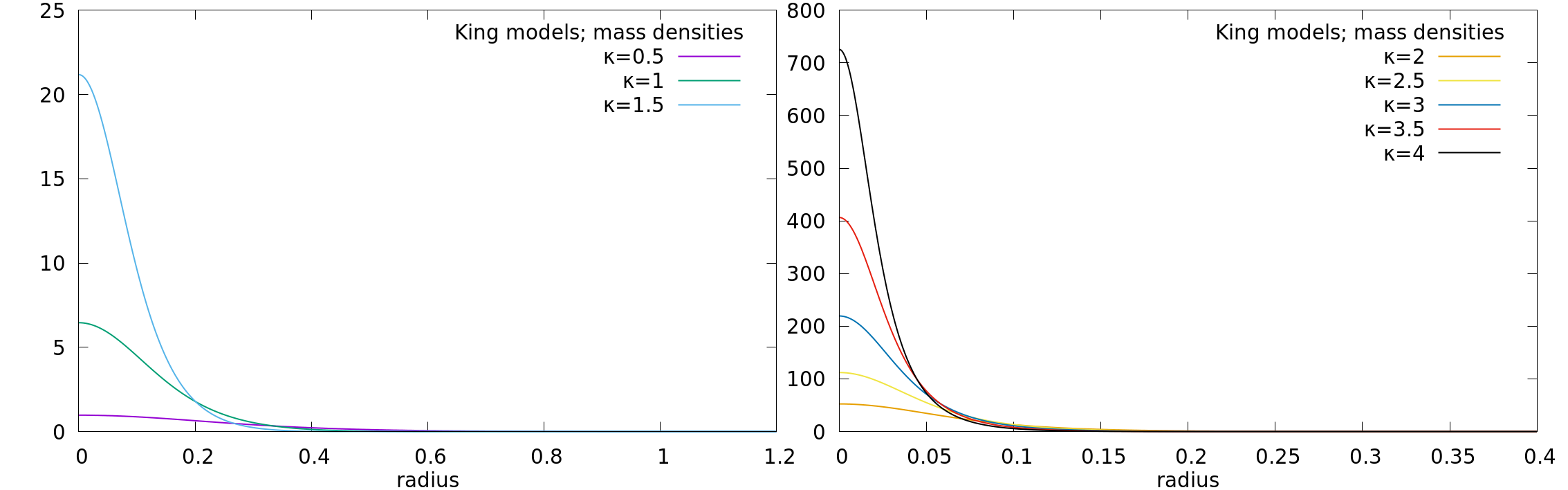}
	\end{center}
	\vspace*{-1.5em}
	\caption{Mass densities~$\rho_0$ of King models~\eqref{eq:f0King} with $\kappa\in\{\frac12,1,\frac32\}$ (left panel) and $\kappa\in\{2,\frac52,3,\frac72,4\}$ (right panel).}
	\label{fig:King_radials}
\end{figure}

For the King models~\eqref{eq:f0King}, the results from~\cite{RaRe17} imply that a scaling law similar to the polytropic case does not hold -- different values of $\kappa>0$ lead to qualitatively different King models.
By~\eqref{eq:rho0centrekappa}, $\kappa$ determines the central density of the King model.
Due to the exponential law included into the King models' energy dependency function, the central density becomes large for large~$\kappa$ (for instance, $\rho_0(0)\approx726$ for $\kappa=4$).
As $\kappa\to\infty$, the total mass~$M_0$ and the maximal radius~$\Rmax$ have positive limit values~\cite{RaRe17}.
On the other hand, the radial extend of the King models becomes larger and larger as $\kappa\to0$.
To be able to accurately handle the King models from a numerics point of view, we will focus on $\kappa$-values in the range $[\frac12,4]$.
The mass densities of a few such King models are shown in Figure~\ref{fig:King_radials}.

For polytropic shells~\eqref{eq:f0polyshell}, the parameter~$\kappa$, together with~$L_0>0$, determines the size of the inner vacuum region via~\eqref{eq:Rminkappa}.
As already done in Figure~\ref{fig:polyshell_radials}, we will fix $\kappa=1$ and focus more on the influence of different values of~$L_0>0$ in this situation.

\section{Radial particle periods in steady states}\label{sc:T}

\subsection{Mathematical background on the radial particle periods}\label{ssc:Ttheory}

Let~$f_0$ be a fixed steady state constructed as above with associated potential~$U_0$.
Using the radial variables~\eqref{eq:defrwL}, the characteristic system~\eqref{eq:charsysxv} determining the particle motions within the steady state takes on the form
\begin{equation}\label{eq:charsysrw}
	\dot r=w,\qquad\dot w=-\Psi_L'(r),
\end{equation}
where~$\Psi_L$ is the effective potential of the steady state:
\begin{equation}
	\Psi_L\colon{]0,\infty[}\to\R,\;\Psi_L(r)\coloneqq U_0(r)+\frac L{2r^2},\qquad L>0.
\end{equation}
The squared modulus of the angular momentum~$L$ can be interpreted as a parameter of the planar ODE~\eqref{eq:charsysrw}.
The particle energy~\eqref{eq:defE}, which can be written as $E(x,v)=E(r,w,L)=\frac12w^2+\Psi_L(r)$, takes on the role of the Hamiltonian function of~\eqref{eq:charsysrw}.

The solutions of~\eqref{eq:charsysrw} are determined by the properties of the effective potential~$\Psi_L$.
For any $L>0$, it is straight-forward to show that there exists a unique radius $r_L>0$ such that $0>\min(\Psi_L)=\Psi_L(r_L)\eqqcolon\Emin L$, cf.~\cite[Lemma~2.1]{HaReSt21} or~\cite[App.~A.1]{Ku21}.
Moreover, for any $E\in{]\Emin L,0[}$, there exist two unique radii $r_-(E,L)<r_L<r_+(E,L)$ such that $\Psi_L(r_\pm(E,L))=E$.
This means that the radial component of any solution of~\eqref{eq:charsysrw} with conserved energy value $E\in{]\Emin L,0[}$ oscillates between the pericenter $r_-(E,L)$ and the apocenter $r_+(E,L)$. 
Let $T(E,L)>0$ denote the period of this motion. 
The induced function $(E,L)\mapsto T(E,L)$ is called the {\em (radial) period function}.
For more background, like the derivation of an integral formula for $T(E,L)$, we refer to~\cite[Sc.~3.1]{BiTr}.

Here we are interested in particles inside the steady state support.
The interior of the set of $(E,L)$-values corresponding to the steady state support is
\begin{equation}\label{eq:defDEL}
	\DEL_0\coloneqq\left\{(E,L)\in{]-\infty,0[}\times{]0,\infty[}\mid L>L_0\,\land\,\Emin L<E<E_0\right\}.
\end{equation}
We refer to this set as the {\em $(E,L)$-support} of the steady state.
Visualisations of~$\DEL_0$ can, e.g., be found in~\cite[Fig.~1.1]{Ku21} or~\cite[Fig.~1]{MoRiBo23} (in the case of isotropic steady states).

The properties of $T\colon\DEL_0\to{]0,\infty[}$ for the steady states introduced in the previous section have been studied in detail in~\cite[App.~B]{HaReSt21}, \cite[Ch.~3]{Ku21}, \cite[App.~B]{RiSa2020}, and~\cite[App.~A]{St24}.
For all of these steady states, the radial period function~$T$ is continuously differentiable on~$\DEL_0$.
For isotropic steady states like~\eqref{eq:f0isopoly}--\eqref{eq:f0King} and shells~\eqref{eq:f0polyshell}, the period function can be continuously extended onto the boundary of~$\DEL_0$, cf.~\cite[Thm.~3.13]{Ku21} and~\cite[Cor.~A.4.2]{St24}.
Thus, it is bounded on~$\DEL_0$ and can also be shown to be bounded away from zero, cf.~\cite[App.~B]{HaReSt21}.
More precisely,
\begin{equation}\label{eq:defTminmax}
	\Tmin\coloneqq\inf_{\DEL_0}(T)>0,\qquad\Tmax\coloneqq\sup_{\DEL_0}(T)<\infty.
\end{equation}
The same statements also hold for the anisotropic polytropes~\eqref{eq:f0poly} with the exception that $T(E,L)$ becomes infinite for $(E,L)=(U_0(0),0)\in\overline\DEL_0$~\cite[Rem.~A.4.5]{St24}, and thus $\Tmax=\infty$.

As mentioned in the introduction, understanding the dynamics close to the fixed steady state relies on additional properties of the radial period function which are yet unproven.
Let us review some examples for this.
In~\cite[Thm.~8.15]{HaReSt21} it is proven that there exist oscillatory modes around polytropic shells~\eqref{eq:f0polyshell} with certain exponents~$k$ and~$\ell$ provided that
\begin{equation}\label{eq:Tmaxcondition}
	\Tmax=T(E_0,L_0).
\end{equation}
This condition is quite natural: It means that the longest radial period~$\Tmax$ within the steady state support corresponds to the particles with the largest radial orbit; notice that $r_-(E_0,L_0)<r_-(E,L)$ and $r_+(E,L)<r_+(E_0,L_0)$ for $(E,L)\in\DEL_0$ by~\cite[Lemma~A.4]{Ku21}.
In~\cite[Cor.~4.16]{Ku21} it is shown that if the period function attains its maximum on~$\overline\DEL_0$ inside the (open) set~$\DEL_0$, there exists an oscillatory mode around the steady state.
Although this result is only stated in~\cite{Ku21} in the case of isotropic steady states, it also holds for the shells~\eqref{eq:f0polyshell}, cf.~\cite[Cor.~5.4.3]{St24}.
In~\cite[Cor.~1.3]{MoRiBo23} it is proven that the number of oscillatory modes with periods larger than all radial particle periods is finite around isotropic polytropes~\eqref{eq:f0isopoly} provided that~\eqref{eq:Tmaxcondition} holds and that the maximum is strict; see also~\cite[Cor.~5.4.4]{St24}.
Further criteria involving the period function as well as the behaviour of other quantities are, e.g., derived in~\cite[Ch.~4]{Ku21} and~\cite[Thm.~1.2(2)]{MoRiBo23}.

The most straight-forward way to establish the condition~\eqref{eq:Tmaxcondition} would be to show that the period function is monotonic on~$\DEL_0$ with respect to both of its variables.
The period function being increasing in~$E$ would also be helpful for other arguments regarding the dynamics close to the steady state, cf.~\cite[Rem.~6.5.6]{St24}.
The only monotonicity results of the period function that have yet been proven for the steady states considered here are those in~\cite[Lemmas~3.14 and~3.15]{Ku21}.
There it is shown that for isotropic steady states, the following properties hold for (the extension of) the radial period on the boundary of~$\DEL_0$:
the functions $L\mapsto T(\Emin L,L)$ and $E\mapsto T(E,0)$ are both strictly increasing on the steady state support.
It is, however, possible to derive further monotonicity results in other settings, e.g., in suitable perturbative regimes, cf.~\cite[Lemma~3.6]{HaReScSt23} or~\cite[Prop.~6.2.6]{St24}, or in other symmetry classes, cf.~\cite[Prop.~2.7]{HaReSt21}.
Furthermore, there is quite a rich literature studying period functions in more general settings and deriving criteria for the periods to be monotonic, cf., e.g., \cite{Ch85,ChKe96,ChWa86,Sc85}.
However, these general criteria could not yet be verified in the present setting, see~\cite[App.~A.3.3]{St24} for a more detailed discussion.

\subsection{Numerical observations regarding the radial particle periods}\label{ssc:Tnum}

\begin{figure}[h!]	
	\begin{center}
		\centering
		\includegraphics[width=\textwidth]{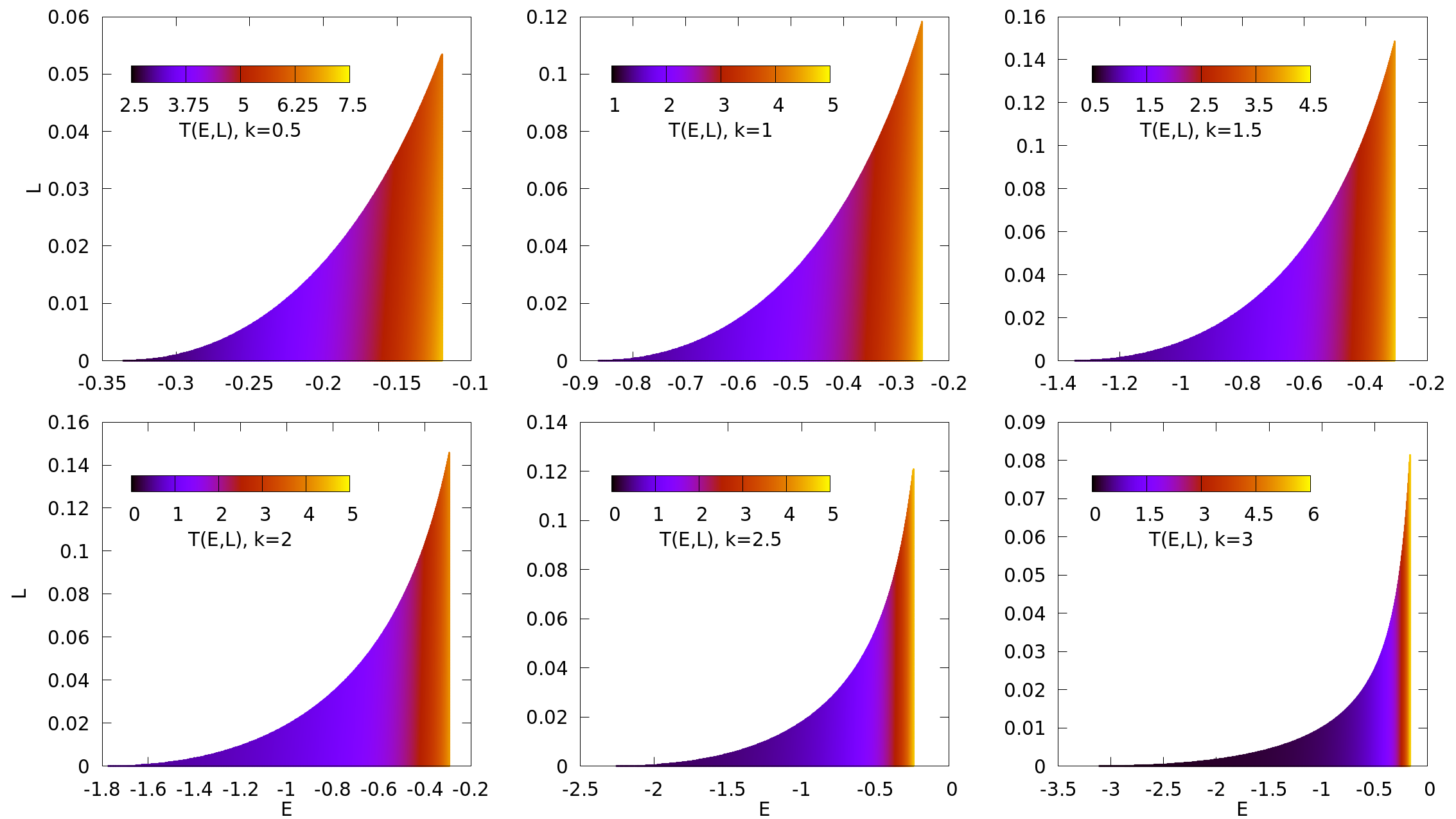}
	\end{center}
	\vspace*{-1.5em}
	\caption{The $(E,L)$-supports~$\DEL_0$ of isotropic polytropes~\eqref{eq:f0isopoly} with polytropic exponents $k\in\{\frac12,1,\frac32,2,\frac52,3\}$ and $\Rmax=1$ together with the values of the associated period functions~$T$ (colour gradient) attained on them.}
	\label{fig:isopoly_periods}
\end{figure}

Let us now numerically analyse the period functions on the $(E,L)$-supports~\eqref{eq:defDEL} of the steady states from Section~\ref{sc:stst} regarding the properties motivated above.
We start with the isotropic polytropes~\eqref{eq:f0isopoly} with different polytropic exponents~$k$, cf.\ Figure~\ref{fig:isopoly_periods}; as discussed in Section~\ref{sc:stst}, we always choose the parameter $\kappa>0$ such that $\Rmax=1$ and only consider $0<k\leq3$.
For the isotropic polytropes included in Figure~\ref{fig:isopoly_periods} as well as for further polytropic exponents~$k$, we observe the following properties of the period function~$T$:
The maximum of (the continuous extension of)~$T$ on~$\overline\DEL_0$ is always attained at $(E_0,0)$; this point corresponds to the bottom right corner in each of the plots in Figure~\ref{fig:isopoly_periods}.
This is always the only point where the maximum is attained.
Furthermore, as indicated by Figure~\ref{fig:isopoly_periods}, the period function is always increasing in~$E$ on~$\DEL_0$, i.e., $T(\cdot,L)$ is increasing on $]\Emin L,E_0[$ for any $0<L<\Lmax\coloneqq\sup\{\tilde L>0\mid \Emin{\tilde L}<E_0\}$.
We also discovered a promising approach to prove this monotonicity rigorously:
The $E$-derivate of~$T$ possesses the integral representation
\begin{equation}\label{eq:TpartialEint}
	\partial_ET(E,L)=\frac1{E-\Emin L}\int_{r_-(E,L)}^{r_L}\frac{\Gref_L(r)}{\sqrt{2E-2\Psi_L(r)}}\diff r,\qquad(E,L)\in\DEL_0,
\end{equation}
where $\Gref_L\colon{]r_-(E,L),r_L[}\to\R$ consists of radial derivatives of the effective potential~$\Psi_L$ up to second order as well as a suitable reflection type mapping ${]r_-(E,L),r_L[}\to{]r_L,r_+(E,L)[}$, cf.~\cite[Lemma~A.3.22]{St24}.
Numerical computation of~$\Gref_L$ reveals that it is always positive on the entire domain $]r_-(E,L),r_L[$ for any $0<L<\Lmax$, which fits to the $E$-monotonicity of~$T$ discussed above.
The $L$-monotonicity of~$T$ for the isotropic polytropes is more diverse: 
For not too large values of the polytropic exponent $k>0$, up to about $2.35$, the period function is decreasing in~$L$ on~$\DEL_0$.
For larger polytropic exponents, however,~$T$ is not monotonic in~$L$.

For the King models~\eqref{eq:f0King} we find that the period function behaves similarly to the case of the isotropic polytropes:
The maximum of~$T$ on~$\overline\DEL_0$ is always attained at $(E_0,0)$, this maximum is strict, and the period function is always increasing in~$E$ on~$\DEL_0$.
Again, the function~$\Gref_L$ appearing in the integral representation~\eqref{eq:TpartialEint} is always positive on the entire domain of integration. 
As discussed in Section~\ref{sc:stst}, we focus on King models with $\kappa\in[\frac12,4]$ here.
For these models, we observe that~$T$ is always decreasing in~$L$ on~$\DEL_0$.
However, after analysing some King models with larger values of~$\kappa$, we suspect that the latter might no longer be the case if one would increase the value of~$\kappa$ further; the other properties seem to hold also for general~$\kappa$.

The same properties of the period function can also be observed for polytropic shells~\eqref{eq:f0polyshell} if the exponents~$k,\ell$ satisfy $k+\ell\leq0$. 
In particular, the condition~\eqref{eq:Tmaxcondition} is satisfied.
We have explicitly tested this for the steady states with polytropic exponents $(k,\ell)\in\{(\frac1{10},-\frac25),(\frac1{10},-\frac14),(\frac1{10},-\frac1{10}), (\frac14,-\frac25),(\frac14,-\frac14),(\frac25,-\frac25)\}$ as well as parameters $\kappa=1$ and $L_0\in\left\{10^{j/2}\mid j\in\{-10,\ldots,2\}\right\}$, but expect it to hold for general choices of the parameters.

\begin{observation}\label{obs:Tmaxcorner}
	For all isotropic steady states we have analysed, the period function~$T$ attains its strict maximum on~$\overline\DEL_0$ at $(E_0,0)$ and is increasing in~$E$ on~$\DEL_0$.
	The same also holds for polytropic shells~\eqref{eq:f0polyshell} with $k+\ell\leq0$.
\end{observation}

These properties do, however, not hold for all steady states from Section~\ref{sc:stst}.
As mentioned above, in the case of an anisotropic polytrope~\eqref{eq:f0poly}, the period function becomes infinite at $(U_0(0),0)\in\partial\DEL_0$. 
By approximating these steady states, we find that the period functions associated to polytropic shells~\eqref{eq:f0polyshell} with $\ell>0$ and $0<L_0\ll1$ attain their maximum on~$\overline\DEL_0$ at $(\Emin{L_0},L_0)$. 
In particular, for these polytropic shells, the condition~\eqref{eq:Tmaxcondition} is not satisfied and~$T$ is not monotonic in~$E$ on~$\DEL_0$.

Nonetheless, there are two further properties of the period function we have observed for all steady states during our analysis:

\begin{observation}\label{obs:Tmaxboundary}
	For all steady states we have analysed, the maximum of the period function~$T$ on~$\overline\DEL_0$ is attained at the boundary of~$\overline\DEL_0$. 
	Furthermore, we have never encountered a situation where~$T$ is constant on a subset of~$\DEL_0$ with positive measure, i.e., the level sets of the period function on~$\DEL_0$ are sets of measure zero.
\end{observation}

\section{Dynamics around steady states}\label{sc:dyn}

\subsection{Mathematical background on the linearised Vlasov-Poisson system}\label{ssc:LVPtheory}

Let~$f_0$ be a fixed steady state as derived in Section~\ref{sc:stst}.
To study the influence of spherically symmetric perturbations on this steady state, we plug the (formal) expression $f_0+\epsilon f+\mathcal O(\epsilon^2)$ with $0<\epsilon\ll1$ into the Vlasov-Poisson system~\eqref{eq:vlasov}--\eqref{eq:rho}.
The function~$f=f(t,x,v)$ describes the perturbation to linear order.
We require~$f$ to be spherically symmetric in the sense of~\eqref{eq:defsphsymm} and to vanish on $\{f_0=0\}$ so that $|\epsilon f|\ll f_0$; see~\cite{BaMoRe95} for a discussion of the latter assumption.
Linearising, i.e., dispensing with terms of order $\mathcal O(\epsilon^2)$, yields the following equation for the evolution of~$f$:
\begin{equation}\label{eq:LVP}
	\partial_tf+\T f-\partial_vf_0\cdot\partial_xU_f=0,
\end{equation}
where~$\T$ is the {\em transport operator} associated to the steady state given by
\begin{equation}\label{eq:deftransport}
	\T=v\cdot\partial_x-\partial_xU_0(x)\cdot\partial_v=\{E,\cdot\}
\end{equation}
and~$U_f$ is the gravitational potential determined by~$f$ via the Poisson equation~\eqref{eq:poisson}--\eqref{eq:rho}.
This system is called the {\em linearised Vlasov-Poisson system}.

In most theoretical studies, this system is transformed in an equivalent second-order system~\cite{HaReSt21,Ku21}.
This is achieved by applying a trick due to Antonov~\cite{An60,IpTh68} and deriving an equation for the odd-in-$v$ part of the linear perturbation $f_-(t,x,v)=\frac12(f(t,x,v)-f(t,x,-v))$, which is of the form
\begin{equation}\label{eq:LVP2order}
	\partial_t^2f_--\L f_-=0.
\end{equation}
This is the {\em second-order formulation of the linearised Vlasov-Poisson system} with {\em linearised operator~$\L$} given by
\begin{equation}\label{eq:deflinop}
	\L\coloneqq-\T^2-\Ri.
\end{equation}
The first term of this operator contains the influence of the steady state flow onto the perturbation via the transport operator~\eqref{eq:deftransport}, while the second term describes the gravitational response of the perturbation and is given by
\begin{equation}\label{eq:response}
	\Ri g\coloneqq-\partial_vf_0\cdot\partial_xU_{\T g},\qquad g=g(x,v).
\end{equation}

For every steady state from Section~\ref{sc:stst},\footnote{In fact, the anisotropic polytropes~\eqref{eq:f0poly} are not included in any of the theoretical works known to the author, but it is straight-forward to extend the results to this class of steady states too.} the linearised operator~$\L$ can be realised as a self-adjoint operator on a suitable subspace of the odd-in-$v$ parts of spherically symmetric perturbations, cf.~\cite[Lemma~4.5]{HaReSt21} or~\cite[Lemma~B.17]{Ku21}.
In addition, by Antonov's coercivity bound~\cite{An62}, the spectrum of~$\L$ is non-negative.
Both of these properties crucially rely on the validity of Antonov's stability criterion~\eqref{eq:varphiprimeneg}.
Furthermore, it can be shown that the essential spectrum of~$\L$, i.e., the spectrum without all isolated eigenvalues of finite multiplicity, is given by the frequencies of the radial particle motions within the steady state, see~\cite[Thm.~5.9]{HaReSt21} or~\cite[Cor.~B.19]{Ku21}.
More precisely,
\begin{equation}\label{eq:Lessspec}
	\sigmaess(\L)=\overline{\left(\frac{2\pi\N}{T(\DEL_0)}\right)^2}=\left(\frac{2\pi\N}{[\Tmin,\Tmax]}\right)^2,
\end{equation}
recall that~$\Tmin$ and~$\Tmax$ are the minimal and maximal radial particle periods within the steady state given by~\eqref{eq:defTminmax}.
The same identity also holds for the absolutely continuous spectrum of~$\L$~\cite[Thm.~1.1]{MoRiBo23}.

The qualitative behaviour of solutions of the linearised Vlasov-Poisson system is determined by the spectral properties of~$\L$.
An undamped oscillatory solution corresponds to a positive eigenvalue of~$\L$~\cite[Sc.~III~e)]{IpTh68}.
In this case, the eigenvalue $\lambda>0$ and the oscillation period~$p$ of the solution are related via
\begin{equation}\label{eq:periodev}
	p=\frac{2\pi}{\sqrt\lambda}.
\end{equation}
If~$\L$ possesses no eigenvalues, the solutions of the linearised Vlasov-Poisson system are expected to be damped at a macroscopic level; see~\cite[Sc.~6]{HaReScSt23} and~\cite[Lemma~C.0.7]{St24} for preliminary results in this direction.

It should be noted that only the Eulerian approach to linearising the Vlasov-Poisson system has been discussed thus far.
In order to analyse the evolution of the phase space support of solutions close to steady states at a linearised level, it is, however, necessary to linearise the system in a Lagrangian way.
Such linearisation schemes are derived in~\cite{HaReSt21,Va83} and lead to an operator equivalent to~\eqref{eq:deflinop}.
For this reason, we limit the discussion to the Eulerian linearisation because it is the one most commonly used in the literature~\cite{An60,BaMoRe95,IpTh68,KaSy85,Ma90}.

\subsection{Numerical observations regarding the linearised Vlasov-Poisson system}\label{ssc:LVPnum}

Let us now numerically study the evolution of solutions of the linearised Vlasov-Poisson system~\eqref{eq:LVP} for different underlying steady states.
We will visualise such evolutions using the following two macroscopic quantities:
\begin{align}
	\Elinkin(f(t))&\coloneqq\frac12\int|v|^2\,f(t,x,v)\diff(x,v),\label{eq:defElinkin}\\
	\Elinpot(f(t))&\coloneqq\int U_0(x)\,f(t,x,v)\diff(x,v).\label{eq:defElinpot}
\end{align}
Their sum
\begin{equation}
	\Elin(f(t))\coloneqq\Elinkin(f(t))+\Elinpot(f(t))=\int E(x,v)\,f(t,x,v)\diff(x,v)\label{eq:defElin}
\end{equation}
is conserved along solutions of the linearised Vlasov-Poisson system; we will always include the evolution of~$\Elin$ in our figures as well to show the accuracy of the numerics.
The quantity~$\Elin$ can be interpreted as the linearisation of the total energy of solutions of the Vlasov-Poisson system~\cite[Sc.~1.5]{Re07};~$\Elinkin$ and~$\Elinpot$ are the kinetic and potential parts of the linearised energy~$\Elin$, respectively.

It should be emphasised that we have also analysed the evolutions of other macroscopic quantities, such as the macroscopic functions~$\rho_f(t)$ or~$U_f(t)$.
However, the observed qualitative behaviours have always been consistent across all macroscopic quantities associated to a solution.

\subsubsection*{Isotropic polytropes}

We start with isotropic polytropic steady states~\eqref{eq:f0isopoly} and analyse the solution of the linearised Vlasov-Poisson system launched by the initial condition
\begin{equation}\label{eq:initialconditionLVP}
	f(0)=w\,\partial_E\varphi(E).
\end{equation} 
Functions of this form naturally appear in the mathematical analysis, see~\cite[Lemma~8.2]{HaReSt21} and~\cite[Ex.~2.1]{Ku21}, which is why we focus on this initial condition here.
Later, we will discuss the case of different initial conditions.

\begin{figure}[h!]	
	\begin{center}
		\centering
		\includegraphics[width=\textwidth]{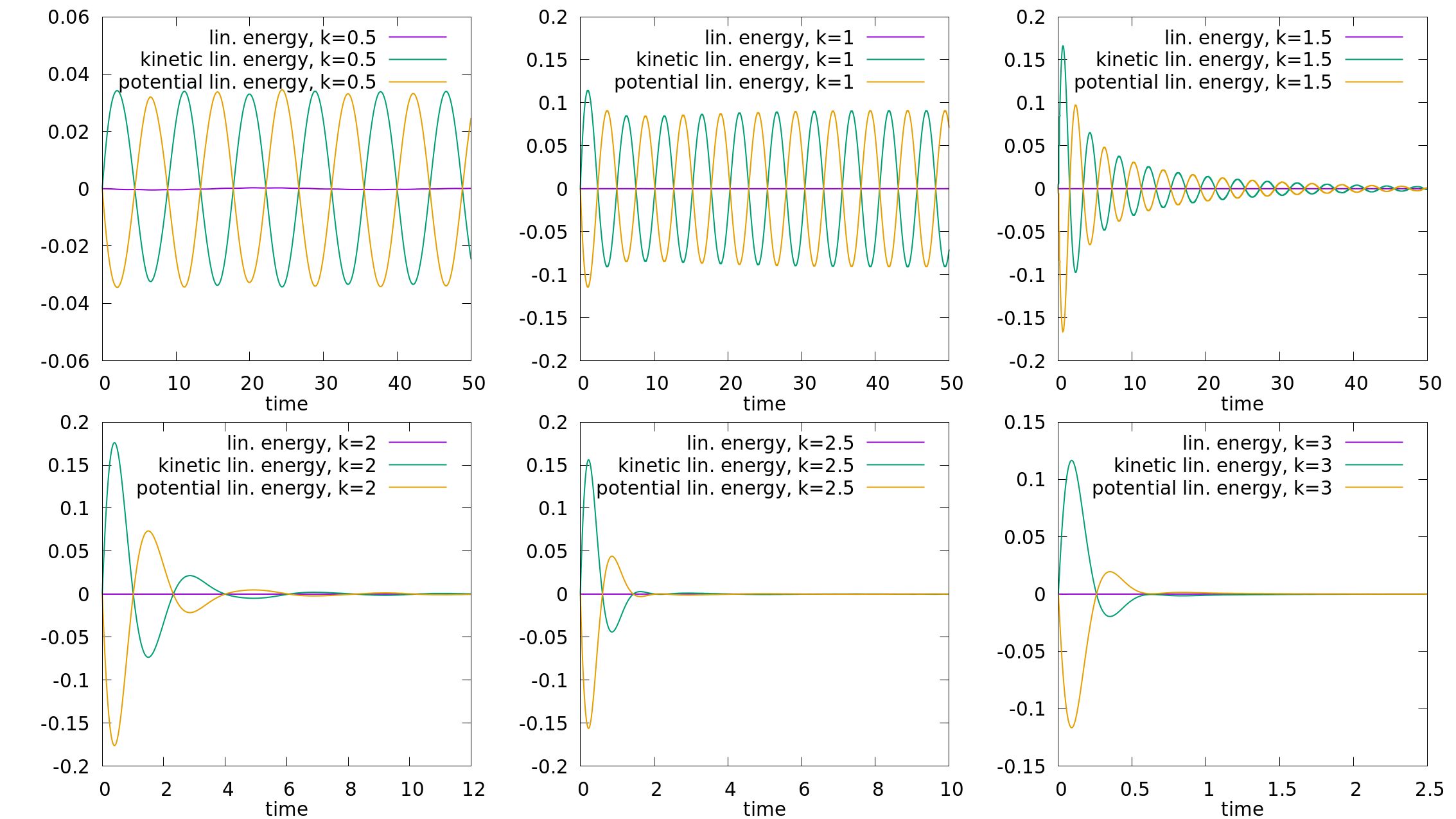}
	\end{center}
	\vspace*{-1.5em}
	\caption{Evolution of the linearised energy~$\Elin$ and its kinetic and potential parts~$\Elinkin$ and~$\Elinpot$ for the solutions of the linearised Vlasov-Poisson systems for isotropic polytropes~\eqref{eq:f0isopoly} with $\Rmax=1$ and polytropic exponents $k\in\{\frac12,1,\frac32,2,\frac52,3\}$. The initial condition is~\eqref{eq:initialconditionLVP} in all cases.}
	\label{fig:isopoly_LVP}
\end{figure}

For the polytropic exponents $k\in\{\frac12,1\}$, the solutions depicted in Figure~\ref{fig:isopoly_LVP} show a (partially) undamped oscillatory behaviour.
More precisely, in the case $k=1$, the solution is partially damped for a brief \enquote{initial damping phase} in the sense that the amplitude decreases during the initial oscillations. Afterwards, the oscillation is undamped.
The same applies to $k=\frac12$, although the initial damping phase is less pronounced.
A qualitatively different behaviour can be observed in the remaining cases of Figure~\ref{fig:isopoly_LVP}:
For $k\in\{\frac32,2,\frac52,3\}$, the solutions again show an oscillatory behaviour, but they are fully damped in the sense that $\Elinkin(f(t))$ and $\Elinpot(f(t))$ decay to zero as~$t$ gets larger.
This damping seems to be stronger, i.e., faster, for larger polytropic exponents~$k$.

\begin{figure}[h!]	
	\begin{center}
		\centering
		\includegraphics[width=\textwidth]{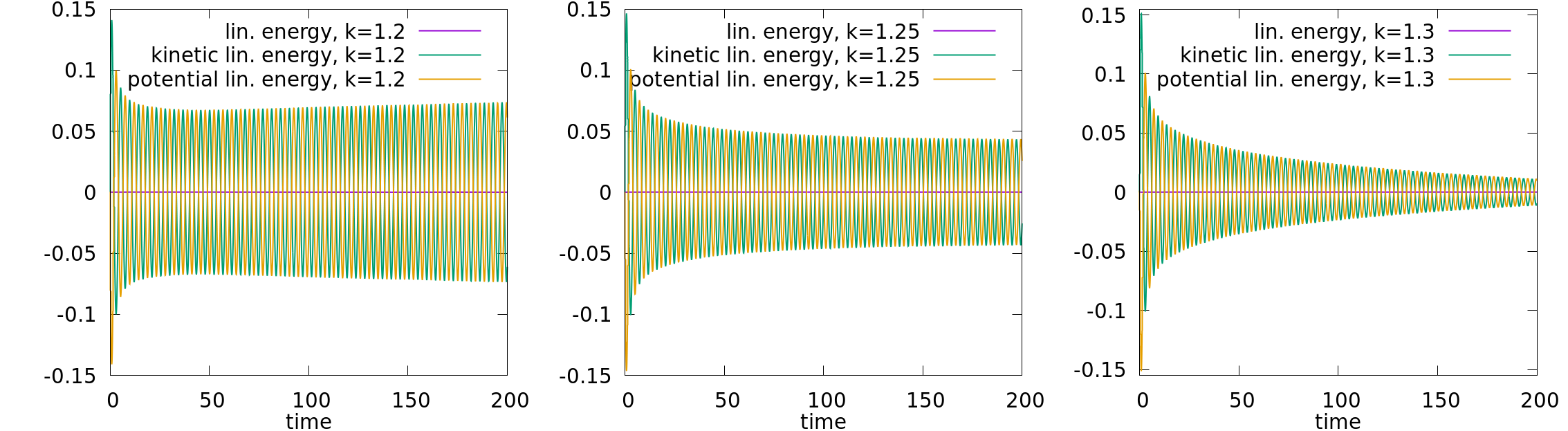}
	\end{center}
	\vspace*{-1.5em}
	\caption{Evolution of~$\Elin$,~$\Elinkin$, and~$\Elinpot$ for the solutions of the linearised Vlasov-Poisson systems for isotropic polytropes~\eqref{eq:f0isopoly} with $\Rmax=1$ and polytropic exponents $k\in\{1.2,1.25,1.3\}$. The initial condition is~\eqref{eq:initialconditionLVP} in all cases.}
	\label{fig:isopoly_trans_LVP}
\end{figure}

In order to determine in more detail the isotropic polytropes for which the solutions of the linearised Vlasov-Poisson system are fully damped and those for which they are not, we analysed similar solutions for more polytropic exponents $0<k\leq3$ than in Figure~\ref{fig:isopoly_LVP}.
The solutions close to the threshold where the qualitative behaviour changes are depicted in Figure~\ref{fig:isopoly_trans_LVP}.
In the case $0<k\leq1.2$, we always observed that the oscillations of the solutions are partially undamped.
The slightly increasing amplitude of the oscillation for $k=1.2$ in Figure~\ref{fig:isopoly_trans_LVP} -- which could be interpreted as a sign of instability of the underlying steady state -- is due to the numerics, which become slightly inaccurate after the large amount of oscillations.
For $1.3\leq k\leq3$, the solutions are always fully damped.
We refrain from assigning the case $k=1.25$ in Figure~\ref{fig:isopoly_trans_LVP} clearly to one behaviour. The solution is probably fully damped in this case too, but the damping is rather slow and there might also be an undamped part of the solution.

Let us now discuss the case of more general initial conditions of the linearised Vlasov-Poisson system.
As explained in Section~\ref{ssc:LVPtheory}, each undamped oscillation corresponds to a positive eigenvalue of~$\L$.
If an initial condition orthogonal to all eigenfunctions to such eigenvalues is chosen, the resulting solution will be fully damped.
However, it is not expected to arrive at such initial condition when choosing it to be of the simple form~\eqref{eq:initialconditionLVP}.
Indeed, we have also analysed solutions of the linearised Vlasov-Poisson system launched by initial conditions different to~\eqref{eq:initialconditionLVP}, e.g., $f(0)=w^jf_0$ and $f(0)=w^j\,\partial_E\varphi(E)$ with $j\in\{0,1,2\}$ as well as $f(0)=\sqrt{r^2+w^2}\,\partial_E\varphi(E)$.
The qualitative behaviours -- i.e., the presence of undamped oscillations and their periods -- were identical to the ones discussed above for the initial condition~\eqref{eq:initialconditionLVP}; the only difference was the extent to which an initial damping phase like in the case $k=1$ in Figure~\ref{fig:isopoly_LVP} was present.
We hence conclude the following:

\begin{observation}\label{obs:LVP_isopoly_oscvdamp}
	For an isotropic polytropic steady state~\eqref{eq:f0isopoly} with polytropic exponent $0<k\leq1.2$, the solutions of the linearised Vlasov-Poisson system (launched by initial data like~\eqref{eq:initialconditionLVP}) oscillate (partially) undamped.
	In contrast, all solutions of the linearised Vlasov-Poisson system are fully damped at a macroscopic level in the case of an isotropic polytrope with $1.3\leq k\leq3$.
\end{observation}

It should be noted that this observation is consistent with previous numerical studies of the dynamics around isotropic polytropes~\eqref{eq:f0isopoly}.
In~\cite{MiSm94}, damped oscillations were found close to the isotropic polytrope with polytropic exponent $k=\frac32$.
These findings were verified in~\cite{WaMu97} based on a numerical method that is more adapted to the Vlasov-Poisson system than the $N$-body code used in~\cite{MiSm94}.
Damped oscillations were also observed for solutions close to the Plummer sphere (which corresponds to an isotropic polytrope with $k=\frac72$) in~\cite{Sw93}.
In~\cite{SePr98} it was found that perturbing isotropic polytropes with exponents $k\leq1$ leads to \enquote{very weakly decaying modes}; we even consider these modes to be undamped.
It is also stated in~\cite{SePr98} that the oscillations are \enquote{strongly damped} for $k\geq1.2$.
Given that the observations from~\cite{SePr98} were derived from $N$-body simulations that had to be conducted with considerably less computational resources than are currently available, this is nonetheless a remarkably high degree of consistency with Observation~\ref{obs:LVP_isopoly_oscvdamp}.
In~\cite{Na00}, solutions close to isotropic polytropes with exponents $k\in\{-\frac32,-\frac12,\frac12,\frac32,\frac52,\frac72\}$ were studied. 
Undamped oscillations were found for $k\leq\frac12$ and a qualitatively different behaviour was identified for $k\geq\frac32$.
The most recent numerical study of the isotropic polytropes was conducted in~\cite{RaRe2018}. 
It is visible in~\cite[Fig.~6]{RaRe2018} that a perturbation of the isotropic polytrope in the case $k=\frac12$ leads to an undamped oscillatory behaviour, that the perturbations get fully damped in the case $k=1.6$, and that the qualitative behaviour is unclear for $k=1.2$.
However, it should be emphasised that all these studies were conducted at the non-linear level, which we will address later in Section~\ref{ssc:VPnum}.

Let us now relate the observed behaviour of solutions of the linearised Vlasov-Poisson system to the theoretical results from the literature~\cite{HaReSt21,Ku21,Ma90,MoRiBo23}.
To do so, we consider the {\em fundamental oscillation period}~$p$ of the oscillatory motion of a solution, i.e., the period of the oscillation which is dominantly visible.
We compute~$p$ by taking the mean distance of all succeeding zeros of $t\mapsto\Elinkin(f(t))$ for the solution launched by the initial condition~\eqref{eq:initialconditionLVP} and multiply this value by~$2$.
In fact, we omit the first few zeros for this computation as the solution's behaviour can be somewhat different at the beginning, e.g., as during the initial damping phase in the case $k=1$ in Figure~\ref{fig:isopoly_LVP}.
Different techniques to determine this period, such as using a discrete Fourier transform, yield qualitatively similar results.

In the situation where the solution contains an undamped oscillatory part, the fundamental oscillation period~$p$ should correspond to an eigenvalue~$\lambda$ of the linearised operator~$\L$; the connection between~$p$ and~$\lambda$ is given by~\eqref{eq:periodev}.
Otherwise, in the case of full damping, the solutions still exhibit periodic oscillations with a reasonably constant fundamental oscillating period~$p$.
In this case, we also convert~$p$ into $\lambda$ using~\eqref{eq:periodev}; $\lambda$ is still expected to be an element of the spectrum of~$\L$, but not an eigenvalue. 
In both cases, we refer to~$\lambda$ as the {\em fundamental spectral element}.

\begin{figure}[h!]	
	\begin{center}
		\centering
		\includegraphics[width=\textwidth]{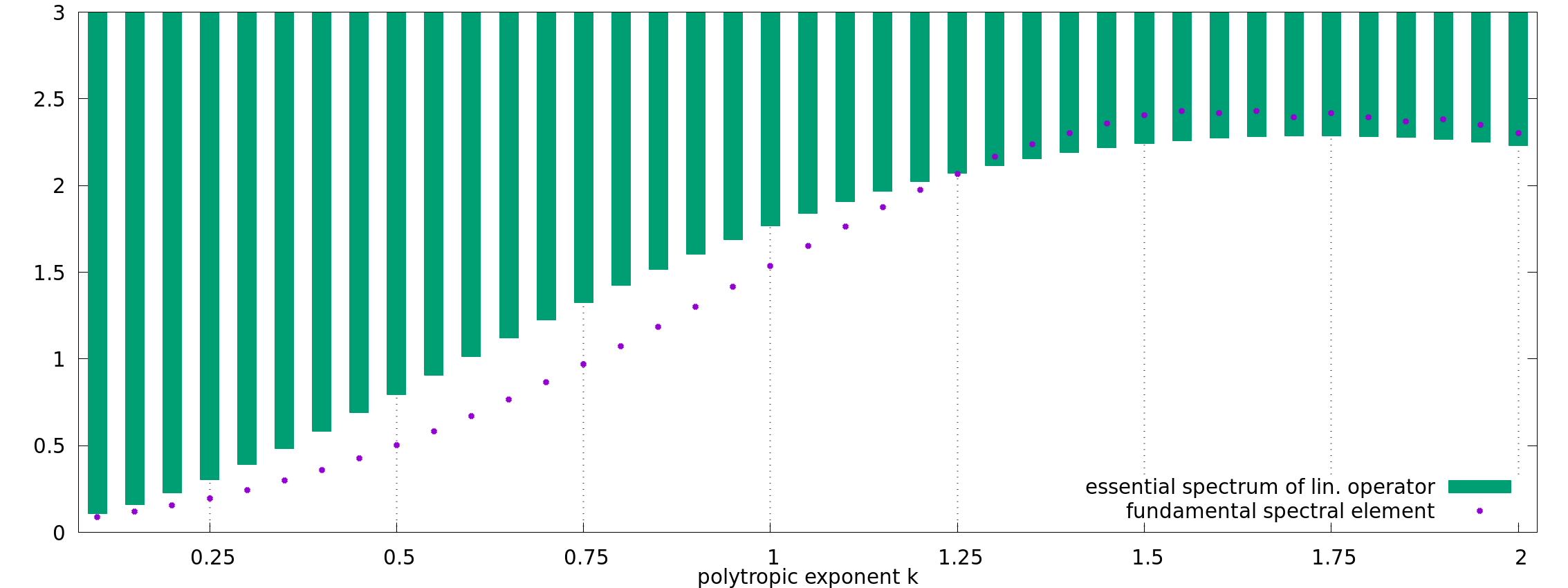}
	\end{center}
	\vspace*{-1.5em}
	\caption{The green boxes illustrate the essential spectra of the linearised operators~$\L$ associated to isotropic polytropes~\eqref{eq:f0isopoly} with $\Rmax=1$ and polytropic exponents $0.1\leq k\leq2$. The purple dots depict the fundamental spectral elements~$\lambda$ which correspond to the fundamental oscillation periods of the solutions of the linearised Vlasov-Poisson system.}
	\label{fig:LVP_spectra_isopoly}
\end{figure}

It is now interesting to compare the fundamental spectral element~$\lambda$ to the essential spectrum of~$\L$.
The reason for this is that in the theoretical works~\cite{HaReSt21,Ku21,Ma90,MoRiBo23}, only the presence of oscillatory modes corresponding to eigenvalues of~$\L$ below the bottom of the essential spectrum is studied. 
Recall that~$\sigmaess(\L)$ is given by~\eqref{eq:Lessspec}, i.e., it is determined by the properties of the radial period function analysed in Section~\ref{sc:T}.
For various isotropic polytropes, this set together with the fundamental spectral element~$\lambda$ is shown in Figure~\ref{fig:LVP_spectra_isopoly}.
What is not visible from the figure is that, for all isotropic polytropes shown, the essential spectrum of~$\L$ is connected.

We see in Figure~\ref{fig:LVP_spectra_isopoly} that the fundamental spectral element~$\lambda$ is always smaller than $\min(\sigmaess(\L))$ for isotropic polytropes with polytropic exponents $0.1\leq k\leq1.2$. 
In the case $k=1.25$, the bottom of the essential spectrum coincides with~$\lambda$ (within the range of the numerical (in)accuracies), while for larger polytropic exponents, $\lambda$ lies inside~$\sigmaess(\L)$.
The similar behaviour persists for all polytropic exponents~$0<k\leq3$ and is not restricted to the range selected in Figure~\ref{fig:LVP_spectra_isopoly} for the sake of better visibility.
We note that this is remarkably consistent with Observation~\ref{obs:LVP_isopoly_oscvdamp}: 
For all isotropic polytropes where the undamped oscillations suggest that the fundamental spectral element~$\lambda$ is an eigenvalue of~$\L$, these eigenvalues are always isolated and not embedded into~$\sigmaess(\L)$.
In other words, the fundamental oscillation period is larger than all individual radial particle periods within the steady state.
In the case $k=\frac12$, this observation has previously been made in~\cite[Fig.~2]{SePr98} at the non-linear level.
For larger polytropic exponents, all solutions of the linearised Vlasov-Poisson system are damped and $\lambda$ always lies in~$\sigmaess(\L)$.
This means that, in every case where the solution is damped, the fundamental oscillation period or an integer multiple of it equals the radial periods of individual particles within the steady state.
It is argued in~\cite[Sc.~5.3]{BiTr} that such resonance necessarily leads to damping.
Let us summarise these findings. 

\begin{observation}\label{obs:LVP_isopoly_noembedded}
	For isotropic polytropic steady states~\eqref{eq:f0isopoly}, there never seems to be an eigenvalue embedded into the essential spectrum of the linearised operator~$\L$.
	More precisely, all eigenvalues seem to lie below the essential spectrum.
	In the fully damped cases, the fundamental oscillation periods correspond to elements of the essential spectrum of~$\L$.
\end{observation}

Let us also mention here that the above observations are consistent with the numerical results obtained in~\cite[Sc.~8.5]{Gue23}, where the linearised Einstein-Vlasov system for isotropic polytropic steady states with small redshift values is investigated.
Such relativistic steady states are close to the respective isotropic polytropes for the Vlasov-Poisson system~\cite{HaRe15} and it is expected that the solutions of the linearised Einstein-Vlasov system behave similarly to the ones of the linearised Vlasov-Poisson system.
It was found in~\cite[Sc.~8.5]{Gue23} that the transition from oscillation to damping occurs close to the polytropic exponent $k=1.2$ for isotropic polytropes with small redshifts.
This is very close to the threshold value observed here, despite the analysis of a different system in~\cite{Gue23} and also despite the fact that the numerical methods used in~\cite[Ch.~8]{Gue23} conceptually differ from the ones employed here.

\subsubsection*{King models}

\begin{figure}[h!]	
	\begin{center}
		\centering
		\includegraphics[width=\textwidth]{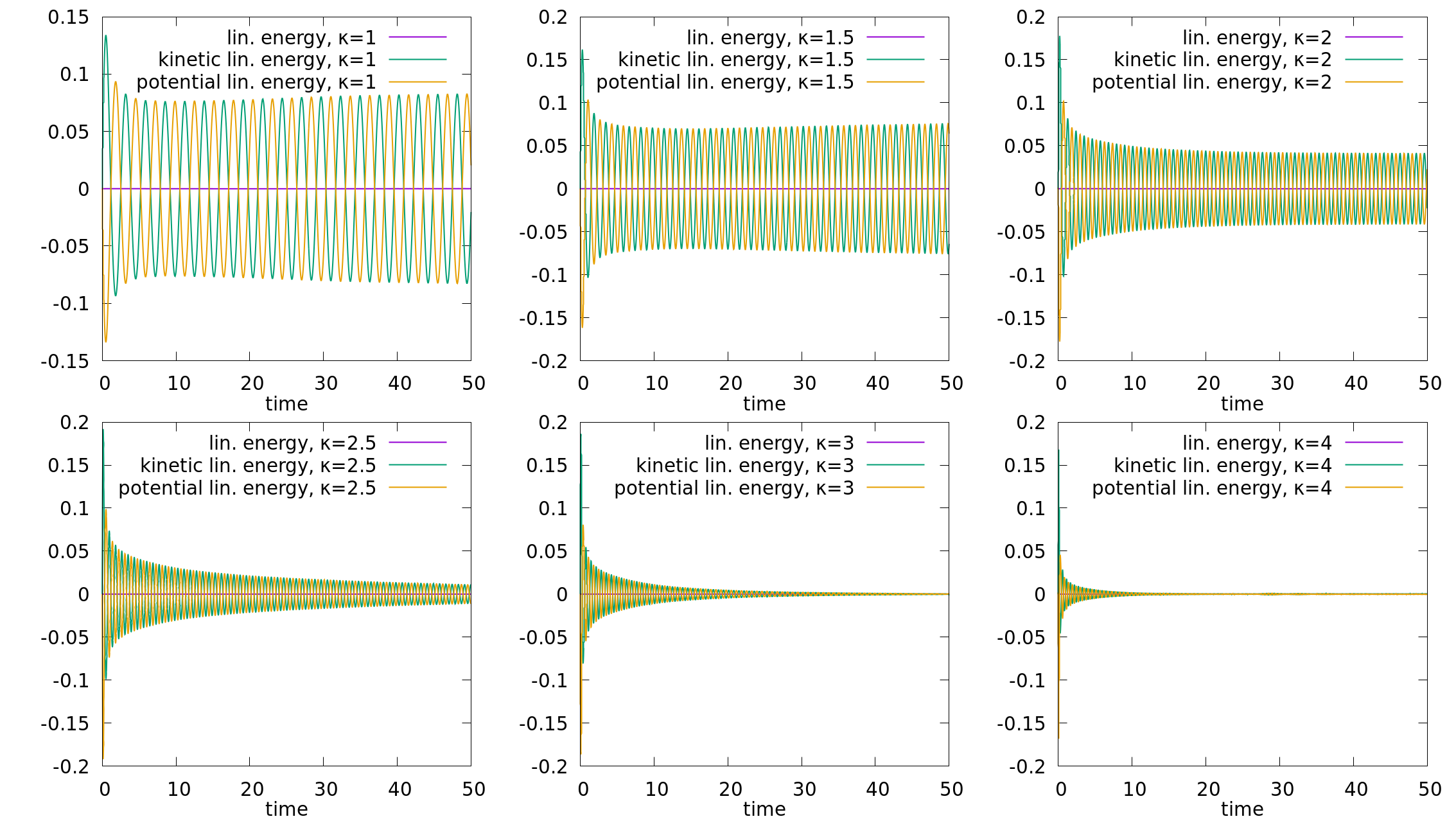}
	\end{center}
	\vspace*{-1.2em}
	\caption{Evolution of~$\Elin$, $\Elinkin$, and~$\Elinpot$ for the solutions of the linearised Vlasov-Poisson systems for King models~\eqref{eq:f0King} with parameters $\kappa\in\{1,\frac32,2,\frac52,3,4\}$. The initial condition is~\eqref{eq:initialconditionLVP} in all cases.}
	\label{fig:LVP_King_few}
\end{figure}

Let us now numerically analyse the situation where the underlying steady state is a King model~\eqref{eq:f0King}.
We keep this part shorter than the above analysis for the isotropic polytropes because the occurring effects are somewhat similar.
The evolution of solutions of the linearised Vlasov-Poisson system for King models with different values of~$\kappa$ are depicted in Figure~\ref{fig:LVP_King_few} in the same way as before.
We have again chosen the initial condition~\eqref{eq:initialconditionLVP}, but note that other initial data lead to similar results.

Figure~\ref{fig:LVP_King_few} shows that King models with small values of~$\kappa$ lead to (partially) undamped oscillatory solutions of the linearised Vlasov-Poisson system, while all solutions are fully damped for larger values of~$\kappa$. 
This also holds for more general $\kappa$-values than those included in the figure.
King models with a large value of~$\kappa$ correspond to more concentrated steady states, recall Figure~\ref{fig:King_radials}.
Because all King models are equally regular, we conclude that the regularity of a steady state -- in particular, at the boundary of its phase space support -- cannot be the only factor determining the existence of undamped oscillatory modes.
It has previously been observed in the literature at the non-linear level that perturbing different King models can either lead to undamped oscillations~\cite{RaRe2018} or to fully damped solutions~\cite[Sc.~2]{SePr98}.

For the King models, it seems rather difficult to determine the precise threshold between undamped oscillations and fully damped behaviour.
Our impression is that this transition proceeds much more slowly for the King models when increasing~$\kappa$ compared to when increasing the polytropic exponent~$k$ for the isotropic polytropes.
In fact, for a wide range of King models, the behaviour of the solutions of the linearised Vlasov-Poisson system is somewhat similar to the case $k=1.25$ in Figure~\ref{fig:isopoly_trans_LVP} which we cannot confidently classify as being fully damped or not.
We are, nonetheless, certain to state the following:

\begin{observation}\label{obs:LVP_King}
	For King models with small values of the parameter~$\kappa$ (at least $\frac12\leq\kappa\leq\frac32$), the solutions of the linearised Vlasov-Poisson system (launched by initial data like~\eqref{eq:initialconditionLVP}) exhibit a partially undamped oscillatory behaviour.
	In contrast, for King models with larger values of~$\kappa$ (at least $\frac52\leq\kappa\leq4$), the solutions of the linearised Vlasov-Poisson system are fully damped (at the macroscopic level).
\end{observation}

In contrast to the isotropic polytropic case, comparing the fundamental spectral elements~$\lambda$ with the essential spectra of the linearised operators~$\L$ does not provide any further insights into the onset of fully damped behaviour for the King models when increasing~$\kappa$.
This is because, for all values of $\kappa>1$ we analysed, the fundamental spectral elements~$\lambda$ are in such close proximity to $\min(\sigmaess(\L))$ that -- given the inherent numerical inaccuracies -- it is not possible to clearly say whether~$\lambda$ lies in the essential spectrum of~$\L$ or just below it.
At the non-linear level, this phenomenon for one fixed King model has previously been observed in~\cite[Fig.~2]{SePr98}.
Let us, nonetheless, emphasise that we have never clearly observed~$\lambda$ lying inside~$\sigmaess(\L)$ in a case where the solutions of the linearised Vlasov-Poisson system partially oscillate undamped or where we are uncertain of the occurring behaviour.
Indeed, for $\frac12\leq\kappa<1$, the fundamental spectral elements always lie below $\min(\sigmaess(\L))$.

\subsubsection*{Anisotropic polytropes}

\begin{figure}[h!]	
	\begin{center}
		\centering
		\includegraphics[width=\textwidth]{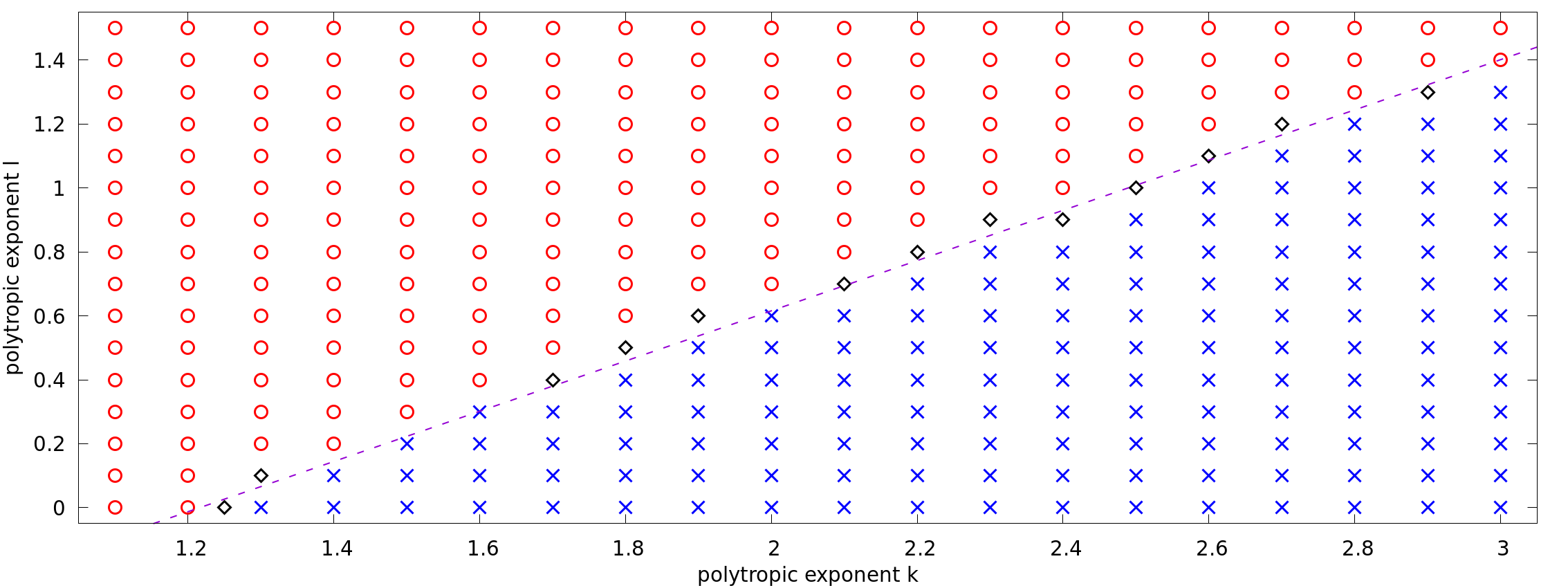}
	\end{center}
	\vspace*{-1.2em}
	\caption{Qualitative behaviour of solutions of the linearised Vlasov-Poisson system for polytropes~\eqref{eq:f0poly} with different polytropic exponents~$k>0$ and $\ell\geq0$ satisfying $k<3\ell+\frac72$. Either there are parts of the solutions which oscillate undamped (red circles), or the solutions are fully damped (blue crosses), or we cannot say for sure which of these two cases is present (black squares). The dashed purple line corresponds to $\frac{\pi^2}{12}k-\frac\pi3\ell=1$.}
	\label{fig:LVP_nonisopoly_oscvdamp}
\end{figure}

We next consider the anisotropic polytropic steady states~\eqref{eq:f0poly}.
As before, we find that the solutions of the linearised Vlasov-Poisson system either oscillate partially undamped or that they are fully damped.
Figure~\ref{fig:LVP_nonisopoly_oscvdamp} displays which polytropic exponents~$k$ and~$\ell$ lead to which of these behaviours. 
We have determined the qualitative behaviour in the same way as above, i.e., by analysing the evolution of~$\Elinkin$ up to a final time of about $t=50$.
In some situations, the behaviour is similar to the isotropic polytropic case $k=1.25$, recall Figure~\ref{fig:isopoly_trans_LVP}, and we cannot confidently assign it clearly to one behaviour.
We used the same initial condition as before, i.e., $f(0)=w\,\partial_E\varphi(E,L)$, but again note that other initial data lead to similar results.
Figure~\ref{fig:LVP_nonisopoly_oscvdamp} can be seen as an extension of~\cite[Fig.~5]{He72}. However, the focus in~\cite{He72} lies on the non-linear (in)stability of anisotropic polytropes with non-positive polytropic exponents~$k$ and~$\ell$, which we do not consider here.

Figure~\ref{fig:LVP_nonisopoly_oscvdamp} shows that undamped oscillatory behaviour exists for smaller values of the polytropic exponent~$k$ and for larger values of the other polytropic exponent~$\ell$. 
Hence, because the smoothness of a polytrope increases with both~$k$ and~$\ell$, cf.~\cite[Rem.~3.17]{Ku21} and~\cite[Rem.~A.3.5]{St24}, we conclude that the presence of an oscillatory mode is not solely determined by the steady state regularity.
Instead, it appears that a greater degree of anisotropy of the steady state, corresponding to a larger value of~$\ell$, increases the likelihood of undamped oscillations.
This has previously been observed in~\cite{RaRe2018} at the non-linear level. There it is argued that steady states with small~$\ell$ are somewhat more homogeneous which, in the light of the results known in the plasma physics case~\cite{MoVi11}, ought to correspond to more damping.

A new insight provided by Figure~\ref{fig:LVP_nonisopoly_oscvdamp} is that the presence of undamped oscillations or damping appears to depend on the polytropic exponents~$k$ and~$\ell$ in a linear way.
Concretely, the dashed line included into the figure -- which has been fitted by manual trial and error -- separates the two regimes quite well, despite the inherent inaccuracies of the numerics.
In the isotropic case $\ell=0$, this line suggests that the transition from undamped oscillations to damping occurs at the polytropic exponent $k=\frac{12}{\pi^2}\approx1.22$, which is consistent with Observation~\ref{obs:LVP_isopoly_oscvdamp}.

\begin{observation}\label{obs:LVP_anisopoly}
	The following linear relation between the polytropic exponents~$k$ and~$\ell$ of an anisotropic polytrope~\eqref{eq:f0poly} and the qualitative behaviour of the solutions of the linearised Vlasov-Poisson system applies to high accuracy:
	There exist undamped oscillatory solutions if and only if
	\begin{equation}
		\frac{\pi^2}{12}\,k-\frac\pi3\,\ell<1.
	\end{equation}
\end{observation}

This observation is consistent with~\cite[Fig.~6~(d)]{RaRe2018}, where an undamped oscillation was found close to a polytrope with $k=3$ and $\ell=5$ at the non-linear level.

Recall that the radial particle periods are unbounded within an anisotropic polytrope with $\ell>0$, cf.\ Section~\ref{ssc:Ttheory}.
Thus, by~\eqref{eq:Lessspec}, the essential spectrum of the linearised operator is of the form $\sigmaess(\L)=[0,\infty[$ in this case.
Because any oscillatory mode corresponds to a positive eigenvalue of~$\L$, the above observation hence shows that we cannot expect the absence of embedded eigenvalues for anisotropic polytropic steady states with $\ell>0$.
This is in sharp contrast to the isotropic case $\ell=0$, recall Observation~\ref{obs:LVP_isopoly_noembedded}.

\subsubsection*{Polytropic shells}

Lastly, we consider polytropic shells~\eqref{eq:f0polyshell} with $\kappa=1$.
As to be expected, if there exist undamped oscillatory solutions for an anisotropic polytrope with exponents $k,\ell>0$, the same can also be observed for polytropic shells with the same polytropic exponents and $0<L_0\ll1$.
In addition, the oscillation periods in these cases are close to each other, as are the values of the radial period functions.
Because $\Tmax=\infty$ in the case $L_0=0$, we also observe that for small values of $L_0>0$, the fundamental oscillation period of an oscillatory solution is in resonance with the radial periods of individual particles.
As discussed above, this corresponds to an eigenvalue of~$\L$ embedded into its essential spectrum.
Hence, the absence of embedded eigenvalues does not hold in general for polytropic shells, although $\Tmax<\infty$ for these models.
For polytropic exponents corresponding to fully damped solutions in the case $L_0=0$, we observe the same behaviour for $0<L_0\ll1$.

When increasing the parameter $L_0>0$, we never observe such fully damped solutions.
For instance, for all polytropic shells with $L_0=1=\kappa$ and polytropic exponents~$k,\ell>0$ contained in Figure~\ref{fig:LVP_nonisopoly_oscvdamp}, the solutions of the linearised Vlasov-Poisson system are oscillating partially undamped.
The fundamental oscillation periods of these solutions correspond to eigenvalues of~$\L$ below the essential spectrum.

The same behaviour of solutions can also be observed for general~$L_0>0$ if the polytropic exponents satisfy $k+\ell\leq0$; we have again analysed the explicit examples listed before Observation~\ref{obs:Tmaxcorner}.
In total, we conclude the following:

\begin{observation}\label{obs:LVP_polyshell}
	For all polytropic shells~\eqref{eq:f0polyshell} with polytropic exponents $k,\ell>0$ and $0<L_0\ll1$, the solutions of the linearised Vlasov-Poisson system behave similarly to the respective solutions in the anisotropic polytropic case $L_0=0$ (cf.\ Observation~\ref{obs:LVP_anisopoly}).
	If~$L_0>0$ is not too small, the solutions of the linearised Vlasov-Poisson system always possess an undamped oscillatory part, regardless of~$k$ and~$\ell$.
	Undamped oscillatory solutions can also be observed for all polytropic shells with general $L_0>0$ provided that $k+\ell\leq0$.
\end{observation}

Perturbations of polytropic shells at the non-linear level have been analysed in~\cite[Sc.~5]{RaRe2018}.
There, undamped oscillations were found close to all polytropic shells; a likely explanation for this is that very small values of~$L_0>0$ were not considered in~\cite{RaRe2018}.

\subsection{Mathematical explanations for the linearised dynamics}\label{sssc:LVPproofs}

Let us now discuss whether the observations from the previous section regarding the solutions of the linearised Vlasov-Poisson system can be explained rigorously by results derived in the literature.

As reviewed in Section~\ref{ssc:Ttheory}, the properties of the radial period function~$T$ can imply the presence of oscillatory modes in certain situations.
For instance, \cite[Thm.~8.15]{HaReSt21} together with the properties of~$T$ from Observation~\ref{obs:Tmaxcorner} prove the existence of oscillatory modes around polytropic shells~\eqref{eq:f0polyshell} with $k+\ell\leq0$.
This is consistent with our simulations of the solutions of the linearised Vlasov-Poisson system in these situations, cf.\ Observation~\ref{obs:LVP_polyshell}.

Another criterion for the existence of undamped oscillatory solutions is derived in~\cite[Cor.~2.2]{Ku21}:
If
\begin{equation}\label{eq:Kunzecrit}
	\sup_{\Rmin<r\leq\Rmax}\frac{U_0'(r)}r<\frac{4\pi^2}{\Tmax^2},
\end{equation}
the linearised operator~$\L$ possesses an eigenvalue below its essential spectrum, i.e., an eigenvalue between~$0$ and $\min(\sigmaess(\L))=\frac{4\pi^2}{\Tmax^2}$.
Although this criterion is only proven in~\cite{Ku21} for suitable isotropic steady states, it can be extended to all steady states from Section~\ref{sc:stst}, cf.~\cite[Lemma~4.5.19]{St24}.
However, for isotropic steady states, one can use that $\Rmin=0$, \eqref{eq:rho0centrekappa}, and the monotonicity of $\frac{U_0'}r$~\cite[Lemma~A.6~(a)]{Ku21} to rewrite the left-hand side of~\eqref{eq:Kunzecrit} as follows:
\begin{equation}
	\sup_{\Rmin<r\leq\Rmax}\frac{U_0'(r)}r=\frac{4\pi}3\,\rho_0(0)=\frac{4\pi}3\,g(\kappa).
\end{equation}
We have checked~\eqref{eq:Kunzecrit} for all steady states analysed above and indeed identified two situations where the criterion is satisfied.
Firstly, \eqref{eq:Kunzecrit} holds for isotropic polytropes~\eqref{eq:f0isopoly} with small polytropic exponents $k>0$ (smaller than about $0.03$).
One way to prove this rigorously could be to consider the limiting case $k=0$.
Although this case is not included in most parts of the analysis because the stability condition~\eqref{eq:varphiprimeneg} does not hold, the corresponding steady state is rather simple ($f_0\equiv1$ inside the steady state support).
It might hence be possible to show the validity of~\eqref{eq:Kunzecrit} in this situation -- concretely, our simulations show that the ratio of the right-hand side and left-hand side of~\eqref{eq:Kunzecrit} is about $1.03$ for the isotropic polytrope~\eqref{eq:f0isopoly} with $k=0$.
Secondly, \eqref{eq:Kunzecrit} is satisfied for polytropic shells~\eqref{eq:f0polyshell} provided that the parameter $L_0>0$ is not too small.
For instance, \eqref{eq:Kunzecrit} holds in the case $k=\ell=\kappa=1$ as long as $L_0\geq\frac1{100}$.
The similar behaviour can also be observed for general polytropic exponents~$k$ and~$\ell$, regardless of whether~$\ell$ is positive or negative.
It should be noted that all these situations where~\eqref{eq:Kunzecrit} implies the existence of oscillatory modes are consistent with the behaviour of solutions of the linearised system observed above, recall Observations~\ref{obs:LVP_isopoly_oscvdamp} and~\ref{obs:LVP_polyshell}.

Another criterion for the presence of oscillatory modes is~\cite[Cor.~4.16]{Ku21}.
However, none of the steady states considered here satisfy it, cf.\ Observation~\ref{obs:Tmaxcorner}.

Other criteria for the existence of undamped oscillatory solutions of the linearised Vlasov-Poisson system like~\cite[Thm.~8.11(a)]{HaReSt21} and~\cite[Thm.~4.13]{Ku21} involve quantities which depend on the steady state in a rather difficult way.
It is, of course, desirable to be able to numerically check these criteria in the future, cf.\ Section~\ref{sc:out}.

The damped behaviour of solutions of the linearised Vlasov-Poisson system is quite elusive from a mathematics point of view; see~\cite{HaReScSt23} for preliminary results in this direction.

\subsection{Comparison between pure transport, linearised, and non-linear system}\label{ssc:VPnum}

In this section we compare the behaviour of solutions of the linearised Vlasov-Poisson system observed above to that of related systems.
Most importantly, we investigate whether solutions of the linearised system indeed explain the behaviour of solutions of the (non-linearised) Vlasov-Poisson system~\eqref{eq:vlasov}--\eqref{eq:rho} close to steady states.
Although solutions of the latter type have often been investigated numerically before~\cite{BaGoHu86,He72,LeCoBi93,MiSm94,Na00,PeAlAlSc96,RaRe2018,SePr98,WaRyMu93,We94}, the transition from non-linear to linearised system has not yet been studied.
It is, however, crucial to understand whether any effects occur during this transition because the behaviour of solutions at the non-linear level is often explained by theoretical considerations at the (mathematically simpler) linearised level.

Similar to Section~\ref{ssc:LVPnum}, we will visualise the behaviour of solutions of the non-linear Vlasov-Poisson system using their kinetic energy
\begin{equation}\label{eq:defEkin}
	\Ekin(f(t))\coloneqq\Elinkin(f(t))=\frac12\int|v|^2\,f(t,x,v)\diff(x,v).
\end{equation}
The simulations in~\cite{RaRe2018} show that the qualitative behaviour of a solution of the non-linear Vlasov-Poisson can be observed alike in the evolution of~$\Ekin$ or in the one of other macroscopic quantities like the functions~$\rho$ and~$U$.
In particular, the qualitative behaviour is also visible in the outer radius of the solution
\begin{equation}
	\Rmax(t)\coloneqq\sup\{|x|\mid\,\exists(x,v)\in\R^3\times\R^3\colon f(t,x,v)\neq0\},
\end{equation}
so that any oscillation is of pulsating nature, i.e., the (radial) support of the solution expands and contracts in a time-periodic way.

\begin{figure}[h!]	
	\begin{center}
		\centering
		\includegraphics[width=\textwidth]{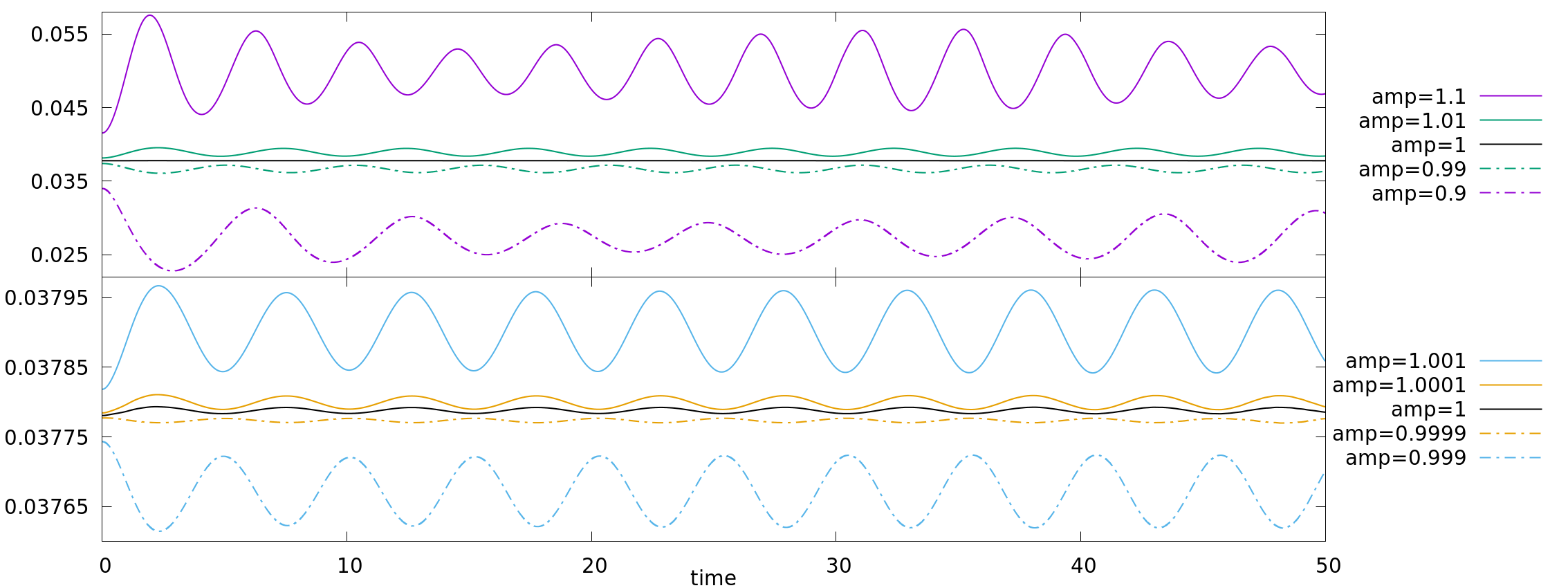}
	\end{center}
	\vspace*{-1.2em}
	\caption{Evolution of the kinetic energy~$\Ekin$ for the solutions of the Vlasov-Poisson system launched by the initial condition~\eqref{eq:pertamp} with perturbation amplitudes $\alpha\in\{1.1,1.01,1.001,1.0001,1,0.9999,0.999,0.99,0.9\}$, where~$f_0$ is the isotropic polytrope~\eqref{eq:f0isopoly} with $k=1$ and $\Rmax=1$.}
	\label{fig:VP_k=1_amps}
\end{figure}

For the non-linear Vlasov-Poisson system, we consider solutions launched by the initial condition
\begin{equation}\label{eq:pertamp}
	f(0)=\alpha\,f_0,
\end{equation}
which should be interpreted as a perturbation of the steady state~$f_0$.
The strength of the perturbation is given by the difference of the {\em perturbation amplitude}~$\alpha$ and~$1$.
In~\cite{RaRe2018} it was found numerically that different types of perturbations -- including physically more natural ones such as dynamically accessible perturbations -- lead to qualitatively similar solutions.
For different choices of~$\alpha$ and a fixed isotropic polytropic steady state, the resulting solutions are shown in Figure~\ref{fig:VP_k=1_amps}.

Because the underlying steady state used in Figure~\ref{fig:VP_k=1_amps} is stable, the solutions remain closer to the steady state if one decreases the strength of the perturbation.
This effect is also present around steady states where the solutions behave in a qualitatively different way.
In Figure~\ref{fig:VP_k=1_amps}, we see that all solutions are undamped oscillatory.
However, the oscillation periods are not identical; for instance, they are (slightly) different for $\alpha=1.01$ and $\alpha=1.0001$.
Nonetheless, the oscillation periods seem to converge to a fixed value as the strengths of the perturbations tend to zero.
The same observation has previously been made in~\cite[Fig.~5]{RaRe2018}; there, also different types of perturbations were taken into account.
Notice that in the case $\alpha=1$, the slight oscillation of the solution is due to numerical errors acting in the same way as a perturbation on the steady state.

As an aside, let us further note that in the case of a strong perturbation ($\alpha=1.1$ or $\alpha=0.9$ in Figure~\ref{fig:VP_k=1_amps}), the solution shows a superposition of multiple oscillatory motions.
An explanation for this might be that the strong perturbation carries parts of the solution to different steady states, around
which they oscillate with differing periods.
Multiple oscillations for rather strong perturbations of steady states have also been observed in~\cite[Figs.~7 and~8]{RaRe2018}.
We will not discuss this phenomenon here further and instead focus on solutions that are closer to a steady state because only these solutions are related to the linearised Vlasov-Poisson system.

\begin{figure}[h!]	
	\begin{center}
		\centering
		\includegraphics[width=\textwidth]{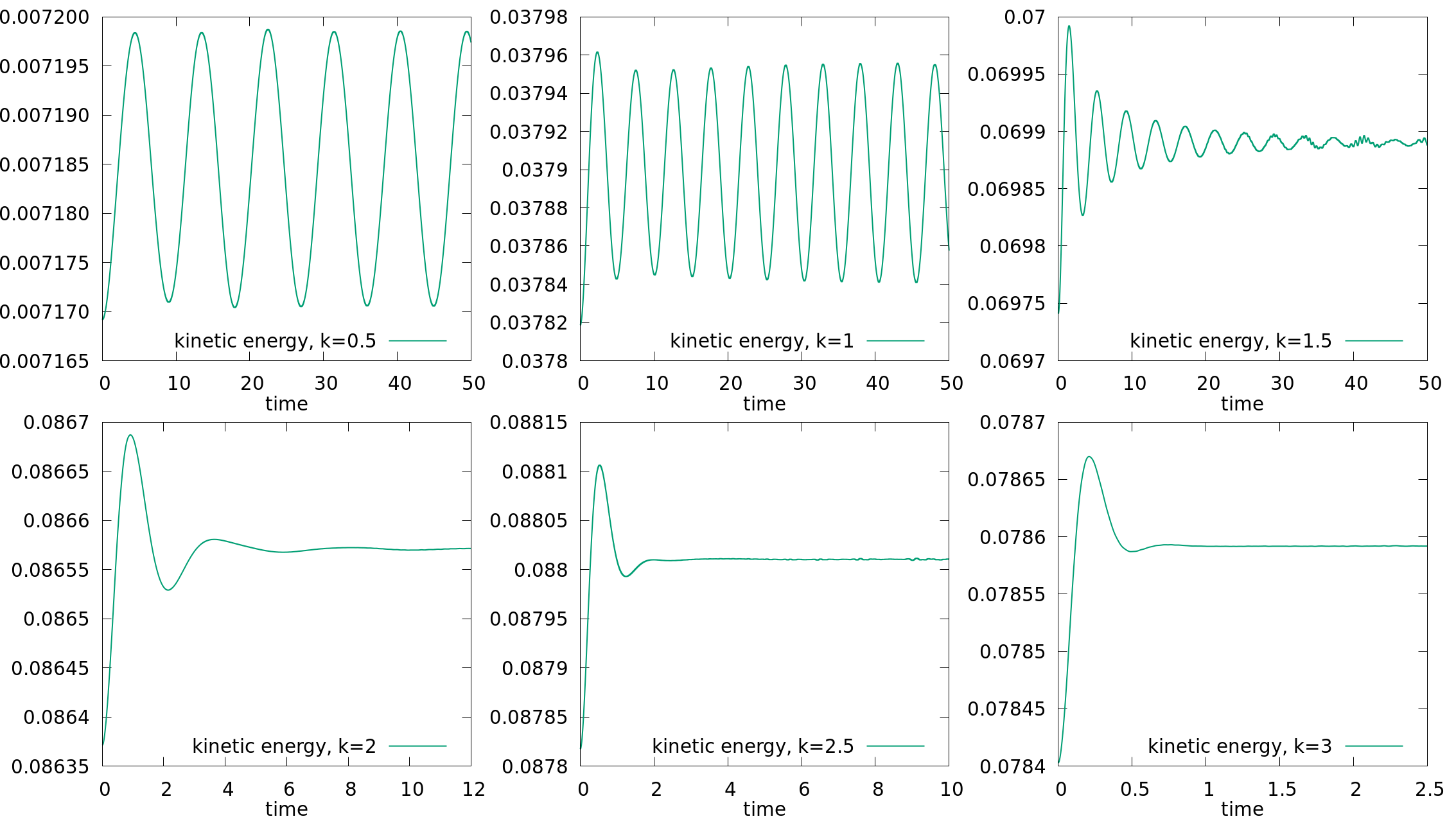}
	\end{center}
	\vspace*{-1.5em}
	\caption{Evolution of the kinetic energy~$\Ekin$ for the solutions of the Vlasov-Poisson system launched by the initial condition~\eqref{eq:pertamp} with $\alpha=1.001$ for the isotropic polytropes~$f_0$~\eqref{eq:f0isopoly} with $\Rmax=1$ and polytropic exponents $k\in\{\frac12,1,\frac32,2,\frac52,3\}$.}
	\label{fig:isopoly_VP}
\end{figure}

In Figure~\ref{fig:isopoly_VP} we depict the evolution of solutions of the non-linear Vlasov-Poisson system close to some isotropic polytropes~\eqref{eq:f0isopoly}.
This figure is the non-linear analogue of Figure~\ref{fig:isopoly_LVP}; we will discuss the connections between both figures below.

\begin{figure}[h!]	
	\begin{center}
		\centering
		\includegraphics[width=\textwidth]{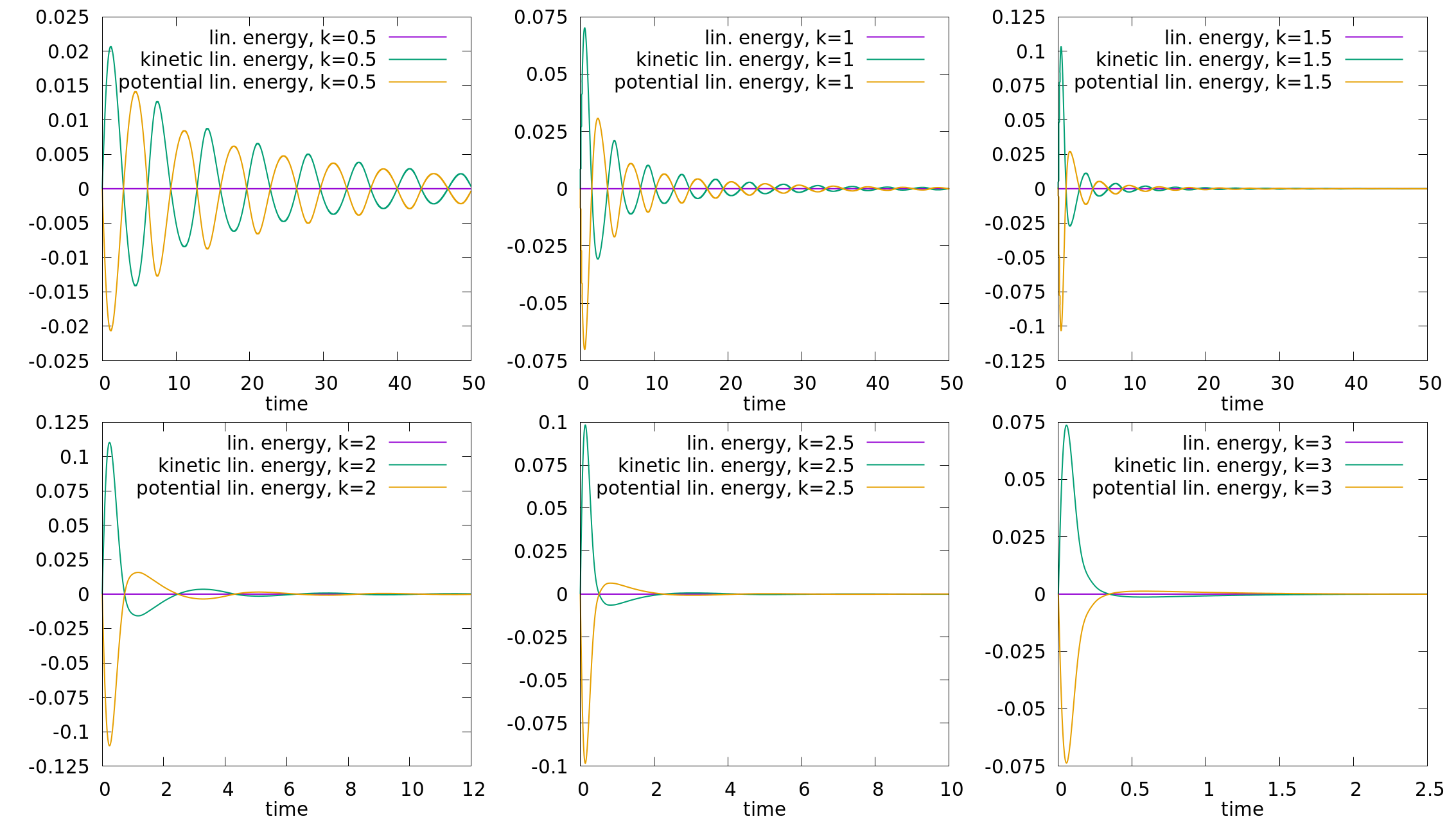}
	\end{center}
	\vspace*{-1.5em}
	\caption{Evolution of the linearised energy~$\Elin$ and its kinetic and potential parts~$\Elinkin$ and~$\Elinpot$ for the solutions of the pure transport equation~\eqref{eq:puretransport} for isotropic polytropes~\eqref{eq:f0isopoly} with $\Rmax=1$ and polytropic exponents $k\in\{\frac12,1,\frac32,2,\frac52,3\}$. The initial condition is~\eqref{eq:initialconditionLVP} in all cases.}
	\label{fig:isopoly_PuTr}
\end{figure}

We also compare the solutions of the linearised Vlasov-Poisson system to the ones of the {\em pure transport equation}
\begin{equation}\label{eq:puretransport}
	\partial_tf+\T f=0;
\end{equation}
recall the definition~\eqref{eq:deftransport} of the transport operator~$\T$ associated to a steady state.
In certain settings, it has been argued that influence of the gravitational response of the perturbation in the linearised Vlasov-Poisson system~\eqref{eq:LVP} is negligible~\cite[p.~280]{LyBe62}, which leads to the simplified system~\eqref{eq:puretransport}.
Figure~\ref{fig:isopoly_PuTr} depicts the solutions of the pure transport equation which are similar to those from Figure~\ref{fig:isopoly_LVP} for the linearised Vlasov-Poisson system.

We can see in Figure~\ref{fig:isopoly_PuTr} that the solutions of the pure transport equation for isotropic polytropes oscillate, but are all fully damped.
This is to be expected in the light of the damping statements established in~\cite{ChLu22,LyBe62,MoRiBo22,RiSa2020}.
All these results rely on certain assumptions on the period function~$T$ (or related functions) which are related to the strict monotonicity of~$T$ with respect to the particle energy, recall Observation~\ref{obs:Tmaxcorner}.
In Figure~\ref{fig:isopoly_PuTr} we again focused on the initial condition~\eqref{eq:initialconditionLVP} used above.
It should, however, be pointed out that all initial data depending only on~$E$ and~$L$ lead to constant-in-time solutions of~\eqref{eq:puretransport} because these functions are in the nullspace of~$\T$, cf.~\cite[Thm.~2.2]{BaFaHo86}.
All other initial data lead to a fully damped solution.
As an aside, we note that the damping rates of solutions of the pure transport equation depend on the initial conditions. 
Nonetheless, the conclusions we draw below about the solutions of the pure transport equation apply to all initial data.

\begin{figure}[h!]	
	\begin{center}
		\centering
		\includegraphics[width=\textwidth]{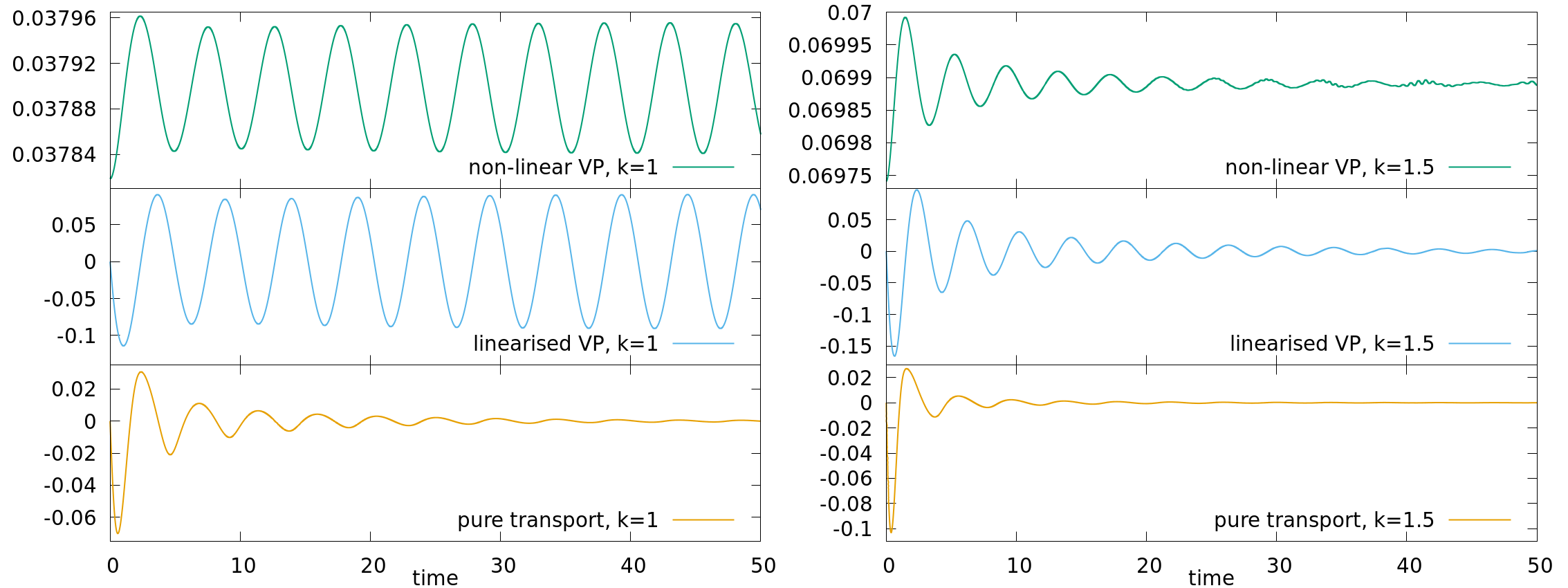}
	\end{center}
	\vspace*{-1.3em}
	\caption{The top panels show the evolutions of the kinetic energy~\eqref{eq:defEkin} for the solutions of the Vlasov-Poisson system launched by the initial condition~\eqref{eq:pertamp} with $\alpha=1.001$ for the isotropic polytropes~$f_0$~\eqref{eq:f0isopoly} with polytropic exponents $k\in\{1,\frac32\}$ and $\Rmax=1$. Below are the evolutions of the linearised kinetic energy~\eqref{eq:defElinkin} of the solutions of the linearised Vlasov-Poisson system (middle) and the pure transport equation~\eqref{eq:puretransport} (bottom) launched by the initial condition~\eqref{eq:initialconditionLVP} for the same two steady states.}
	\label{fig:vgl}
\end{figure}

Comparing Figures~\ref{fig:isopoly_LVP} and~\ref{fig:isopoly_PuTr}, the most striking difference is the presence of undamped oscillations for the linearised Vlasov-Poisson system.
Furthermore, in the cases where the solutions of the linearised Vlasov-Poisson system are fully damped, the damping is stronger for the pure transport equation.
In addition, the fundamental oscillation periods for the solutions of the linearised Vlasov-Poisson system are different to those of the pure transport equation -- the periods are always longer for the former system.
All these differences are visible more clearly in Figure~\ref{fig:vgl}.

Figure~\ref{fig:vgl} shows that the solutions in the isotropic polytropic case $k=1$ at the non-linear and linearised level are both partially undamped oscillatory with a similar oscillation period.
The different behaviours at the beginning of the simulations result in the oscillations to be slightly shifted; this is to be expected because the amplitude perturbation~\eqref{eq:pertamp} for the non-linear system works in a conceptionally different way compared to the initial condition~\eqref{eq:initialconditionLVP} used for the linearised system.
In the case $k=\frac32$, we can see that the fundamental oscillation periods are identical at the non-linear and linearised level and that the damping rates are similar.
The fact that the simulations at the non-linear and at the linear level yield similar results is also visible by comparing Figures~\ref{fig:isopoly_LVP} and~\ref{fig:isopoly_VP}.

So far we have only discussed the connections between non-linear system, linearised system, and the pure transport equation for a few isotropic polytropes.
We have analysed this connection for all steady states considered above, i.e., for more general isotropic polytropes~\eqref{eq:f0isopoly}, King models~\eqref{eq:f0King}, anisotropic polytropes~\eqref{eq:f0poly}, and polytropic shells~\eqref{eq:f0polyshell}.
The findings are always consistent with the above situation, and we thus arrive at the following conclusion:

\begin{observation}\label{obs:VP_lin}
	The solutions of the non-linear Vlasov-Poisson system close to a steady state behave similarly to the solutions of the respective linearised system.
	More precisely, if the solutions of the linearised Vlasov-Poisson system oscillate partially undamped for some steady state, the solutions of the non-linear system close to the same steady state do so as well, and the oscillation periods are similar.
	If the solutions of the linearised system are fully damped at the macroscopic level, so are the solutions of the non-linear Vlasov-Poisson system close to the respective steady state, and the damping is equally strong.
	All these statements require that the solution of the non-linear system is sufficiently close to the respective steady state, i.e., the perturbation is sufficiently weak.
	
	In contrast, the solutions of the pure transport equation~\eqref{eq:puretransport} behave in a qualitatively different way to the solutions of the Vlasov-Poisson system.
\end{observation}

\section{Outlook}\label{sc:out}

Let us now discuss some topics for further research.
We start with open question for which the extension of the numerical methods developed here should prove enlightening.
Firstly, the numerics can be used to study the damping rates of macroscopic quantities. 
These rates can be determined for solutions of the pure transport system~\eqref{eq:puretransport}, where theoretical results on quantitative decay rates were derived in~\cite{ChLu22,MoRiBo22}.
The numerics allow for the determination of the optimal decay rates for different macroscopic quantities and how these depend on the initial data or the underlying steady state.
The decay rates can also be studied in the context of the full linearised Vlasov-Poisson system, where no quantitative damping results are available. 
In order to be able to carry out such an analysis efficiently for a large number of solutions, however, it is first necessary to determine for which steady states there are fully damped solutions of the linearised Vlasov-Poisson system and for which there are undamped oscillatory parts.
A fast predictor for the latter question can, e.g., be implemented numerically by adapting the deep learning based methods from~\cite{StWo24}.
Secondly, the numerics can be used to investigate which steady states violating~\eqref{eq:varphiprimeneg} are unstable. 
Recall that we are considering the system in spherical symmetry here, i.e., we aim to identify steady states that are unstable against spherically symmetric perturbations.
Some evidence that such radial instabilities are considerably harder to find than instabilities against general perturbations is presented in~\cite{DeMe88,MeZa97}.
Some numerical investigations regarding the radial instability of steady states were conducted in~\cite{BaGoHu86,He72}, theoretical results (at the linearised level) were derived in~\cite{GuLi08}.
The only explicitly known examples for radially unstable steady states are those considered in~\cite{WaGuLiZh13}.
It is also of interest to numerically investigate the evolution of an unstable steady state after a perturbation at the non-linear level; see~\cite{Praktikum20} for a similar study in the relativistic setting.
Thirdly, it is of interest to extend the numerical methods developed here so that they also cover related settings studied in the literature, e.g., the case of steady states with unbounded supports, the system with an external potential, or the relativistic Vlasov-Poisson system.

Let us next discuss a few open mathematical problems (in addition to those already mentioned in Section~\ref{sssc:LVPproofs}) that we consider particularly promising after the numerical analyses carried out here.
The above observations suggest that the presence of undamped oscillations (both at the linearised and non-linear level) is monotonic with respect to certain steady state parameters.
For instance, the presence of undamped oscillations seems to be monotonic in the polytropic exponents~$k$ and~$\ell$ of polytropes~\eqref{eq:f0poly} (cf.\ Observations~\ref{obs:LVP_isopoly_oscvdamp} and~\ref{obs:LVP_anisopoly}), in the parameter~$\kappa$ of the King models~\eqref{eq:f0King} (cf.\ Observation~\ref{obs:LVP_King}), and in the parameter~$L_0$ of polytropic shells~\eqref{eq:f0polyshell} (cf.\ Observation~\ref{obs:LVP_polyshell}).
One way of proving these statements rigorously might be via the approach used in~\cite{HaReSt21,Ku21}, where the presence of oscillatory modes is translated into a single number depending on the steady state to be~$>1$.
It is hence possible that this number is already monotonic with respect to the aforementioned steady state parameters; of course, it would also be helpful to first check this claim numerically.
In addition to such monotonicities, it seems promising to determine the qualitative dynamics around steady states in certain limiting case, in particular, in the following two situations. 
On the one hand, Observation~\ref{obs:LVP_King} shows the presence of macroscopic damping around King models in the limit $\kappa\to\infty$; see~\cite{RaRe17} for a careful analysis of this limit and~\cite[Sc.~3.4]{HaLiRe21} for related results in the relativistic case.
On the other hand, Observation~\ref{obs:LVP_polyshell} shows the presence of oscillatory modes around polytropic shells in the limit $L_0\to\infty$; recall that the discussion in Section~\ref{sssc:LVPproofs} shows that the criterion from~\cite[Cor.~2.2]{Ku21} holds for these steady states.
Finally, Observation~\ref{obs:VP_lin} suggests that the qualitative behaviour of solutions of the Vlasov-Poisson system close to a steady state is accurately described by solutions of the respective linearised system, and it would of course be most interesting to verify this rigorously.
A first step in this direction would be to prove that an oscillatory mode leads to undamped oscillations at the non-linear level.
A similar result has been established in the fluid case in~\cite{Ja16}; the approach used there suggests that the different (but equivalent) linearisation schemes developed in~\cite[Sc.~3]{HaReSt21} are useful for this task.

\section{Conclusion}\label{sc:conc}

In this work we have numerically investigated the dynamics of spherical galaxy models close to stable equilibria, more precisely, the behaviour of solutions around stable steady states of the gravitational Vlasov-Poisson system in spherical symmetry.
As reviewed in Section~\ref{sc:intro}, these dynamics have been extensively studied in both the astrophysical and mathematical literature, yet rigorous results remain scarce.

The steady states considered here are among the most commonly analysed in the literature: Isotropic and anisotropic polytropes, King models, and shells. 
These models have been introduced and visualised in Section~\ref{sc:stst}.
One obstacle that has been encountered in numerous theoretical works on the dynamics surrounding these steady states are unknown properties of the radial particle periods within them. 
In Section~\ref{ssc:Tnum}, we have conducted the first numerical analysis of these particle periods. 
Our findings indicate that for all isotropic steady states and some shells, the longest radial particle periods occur for particles with the largest radial orbit and that the periods increase in the particle energy.

A detailed numerical analysis of the dynamics around the steady states has then been conducted in Section~\ref{sc:dyn}.
The initial focus was on the perturbation analysis at the linearised level because this setting is usually considered in theoretical works but rarely in numerical works.
The analysis confirmed that around each steady state, the solutions are either macroscopically damped or parts of the solutions oscillate around the equilibrium.
A new linear relation between these qualitative behaviours and the exponents of polytropic steady states has been identified.
For the King models, undamped oscillations have been found around not too concentrated steady states, while full damping has been found for more concentrated King models.
For shells with a sufficiently large inner vacuum region we have always observed undamped oscillatory solutions.
In Section~\ref{ssc:VPnum}, the dynamics at the linearised level have then been compared with those of the actual non-linear system close to the respective steady states. 
This comparison provides the first numerical evidence that the qualitative behaviour of solutions of the Vlasov-Poisson system close to equilibria is indeed generally governed by the linearised system.

As outlined in Section~\ref{sc:out}, the analysis conducted here opens up a multitude of new questions.
These encompass both further numerical issues as well as theoretical problems that can hopefully be addressed more effectively with the aid of the findings presented here.

\appendix

\section{Numerical methods}\label{sc:num}

In this appendix we describe the numerical methods employed to obtain the above observations.
The methods are based on the ones used in~\cite{AnRe06,Praktikum20,GueStRe21} to simulate the Einstein-Vlasov system and the ones from~\cite{RaRe2018} used for the Vlasov-Poisson system.
The code used for the simulations is written in \CC\ in an object-oriented way.
We have tried to maintain the code as simple as possible; in particular, by avoiding the use of specialised programming libraries.
It is our intention that the code will facilitate further numerical investigations, which is why we made it publicly available via the following link:
\begin{center}
	\href{https://github.com/c-straub/radVP}{https://github.com/c-straub/radVP}
\end{center}
Most of the simulations were run on the supercomputers provided by the Keylab HPC of the University of Bayreuth.

\subsection{Steady states}\label{ssc:numstst}

For a prescribed ansatz of the form~\eqref{eq:varphigeneral} and $\kappa>0$, it is shown in Section~\ref{sc:stst} that computing a steady state (and the macroscopic functions associated to it) corresponds to solving the integro-differential equation~\eqref{eq:ypoisson} with boundary condition~\eqref{eq:yboundary}. 
This is implemented numerically via the midpoint method with a radial step size $\leq10^{-6}$.
The function~$g$ defined in~\eqref{eq:defg} is evaluated by computing the integral via the composite Simpson's rule with $500$ steps. 
In the polytropic cases~\eqref{eq:f0isopoly} and~\eqref{eq:f0poly}, we instead use the explicit formula for~$g$ mentioned in Section~\ref{sc:stst}.
Families of steady states are calculated by executing several calls of the above algorithm in parallel using pthreads.

As a steady state only needs to be computed once for its further analysis, the steady state computation can be carried out with high accuracy.
We have ensured this accuracy by verifying that increasing or decreasing the step size by a factor of~$2$ during the calculation does not result in any significant changes in the resulting steady state.
For steady states with a concentrated region, however, we had to use smaller step sizes to guarantee an accurate computation.
For instance, we used step sizes between $10^{-7}$ and $10^{-8}$ for King models~\eqref{eq:f0King} with $\kappa$-values larger than~$3$.

\subsection{Radial particle periods}\label{ssc:numperiod}

For a fixed steady state~$f_0$ computed as described above, our aim is to numerically calculate the radial period function~$T(E,L)$ for $(E,L)\in\DEL_0$; recall Section~\ref{ssc:Ttheory} for the definitions of these quantities.
We do not use the integral representation of~$T(E,L)$ mentioned in Section~\ref{ssc:Ttheory} for this task because it corresponds to a singular integrand; this integral representation has been used, e.g., in the numerical study~\cite{WaGuLiZh13}.
Instead, we compute (the relevant parts of) the solution $(R,W)(\cdot,E,L)\colon\R\to{]}0,\infty{[}\times\R$ of the radial characteristic system~\eqref{eq:charsysrw} with parameter~$L$ satisfying the initial condition~$(R,W)(0,E,L)=(r_+(E,L),0)$.
We prefer to use the initial condition $(r_+(E,L),0)$ instead of $(r_-(E,L),0)$ here due to numerical reasons: The effective potential~$\Psi_L(r)$ is rather steep at $r\approx0$, which makes the calculation of~$r_-(E,L)$ prone to numerical errors, in particular, for small values of~$L$.
In order to determine~$T(E,L)$, we just have to wait until the velocity component of the solution $(R,W)(\cdot,E,L)$ becomes positive because
\begin{equation}\label{eq:numTEL}
	\frac12\,T(E,L)=\inf\{s\geq0\mid W(s,E,L)>0\}.
\end{equation}

To compute $(R,W)(\cdot,E,L)$, we actually do not solve~\eqref{eq:charsysrw} directly but instead transform (back) to Cartesian coordinates in which the characteristic system has the form~\eqref{eq:charsysxv}.
More precisely, for given~$(r,w,L)$, associated Cartesian coordinates satisfying~\eqref{eq:defrwL} are, e.g., given by
\begin{equation}
	x=(r,0,0),\qquad v=(w,\frac{\sqrt L}r,0).
\end{equation}
After transforming the initial condition $(r_+(E,L),0,L)$ to Cartesian coordinates in this way, we solve~\eqref{eq:charsysxv} until ${x\cdot v}=rw$ becomes positive and obtain the value of~$T(E,L)$ via~\eqref{eq:numTEL}.
The reason why it is more convenient to work in Cartesian coordinates is that it avoids difficulties that would otherwise arise close to the spatial origin $r=0$; these difficulties would be particularly pronounced for small values of~$L$.
In addition, we can include the purely radial case $L=0$ in this computation and also calculate $T(E,0)$; this value is to be understood as the continuous extension of the period function to $L=0$~\cite[Lemma~3.12]{Ku21}.
The price we pay, however, is that the six-dimensional system~\eqref{eq:charsysxv} is numerically more expensive to solve than the planar system~\eqref{eq:charsysrw}. 
To solve~\eqref{eq:charsysxv} numerically, we use the fourth-order Runge-Kutta method with a time step size $\leq10^{-6}$.
More precisely, the heuristic of utilising the same step size for this computation as for the steady state computation has been employed.
This guarantees that the $T$-computation is sufficiently accurate for steady states with a dense region at their centre. 

One way to verify the accuracy of the above numerical computation of the period function is to compare it with the explicit extension formulae to minimal energies $(\Emin L,L)$. 
By~\cite[Lemma~3.8]{Ku21} or~\cite[Eqn.~(B4)]{RiSa2020}, a continuous extension of~$T$ is given by
\begin{equation}\label{eq:Textminenergies}
	T(\Emin L,L)=\frac{2\pi}{\sqrt{\Psi_L''(r_L)}}=\frac{2\pi}{\sqrt{4\pi\rho_0(r_L)+\frac{U_0'(r_L)}{r_L}}},\qquad L>0.
\end{equation}
Extensive testing has shown that this indeed extends the values of $T(E,L)$ for $(E,L)\in\DEL_0$ in a continuous way with high accuracy.
This demonstrates that our numerical computation of the period function $T(E,L)$ works accurately, even in the near circular regime $E\approx\Emin L$ where the solutions of~\eqref{eq:charsysrw} are confined to a small radial interval.
Furthermore, we have ensured the accuracy of the numerics by verifying that changing the time step size with a factor of~$2$ does not significantly influence the computed values.

For several aspects of the analysis in Section~\ref{ssc:Tnum} -- e.g., to accurately compute the supremum of~$T$ on~$\DEL_0$ -- it is necessary to evaluate the period function on large sets of $(E,L)$-pairs.
This is realised by performing several evaluations of the period function in parallel using pthreads once more.
To determine the minimal and maximal periods~$\Tmin$ and~$\Tmax$ as well as to create plots like the ones in Figure~\ref{fig:isopoly_periods}, we computed the period function on a grid containing between $10^5$ and $10^6$ values of $(E,L)\in\DEL_0$.
When computing~$\Tmin$ and~$\Tmax$, we also used the values of the extension~\eqref{eq:Textminenergies} and, in the case $L_0=0$ in~\eqref{eq:varphigeneral}, the extension of~$T$ to $L=0$.
In order to investigate the monotonicity of~$T$ on~$\DEL_0$ with respect to one of its variables, we analysed the functions $T(\cdot,L)$ and $T(E,\cdot)$ for more than $500$ values of~$L$ and~$E$, respectively, and allowed for an error of $5\cdot10^{-5}$.

\subsection{The linearised Vlasov-Poisson system}\label{ssc:numLVP}

For a fixed steady state~$f_0$ computed as described in Section~\ref{ssc:numstst}, let us now describe how to simulate the associated linearised Vlasov-Poisson system; recall Section~\ref{ssc:LVPtheory} for theoretical backgrounds on this system.
We have adapted the numerical method from that used to simulate the non-linear Vlasov-Poisson system, which we will discuss below in Section~\ref{ssc:numVP}.
Later we discovered that similar methods had already been used in the astrophysics literature (albeit in different settings) in~\cite{LeCoBi93,Me87,WaRyMu93}. 

We first rewrite the linearised Vlasov-Poisson system~\eqref{eq:LVP} in a more convenient form for purposes of numerical simulations.
In particular, we do not use second-order formulations like~\eqref{eq:LVP2order} here because first-order formulations are more practical for the numerics. 
For $(x,v)\in\R^3\times\R^3$, let $(X,V)(\cdot,x,v)\colon\R\to\R^3\times\R^3$ denote the solution of the characteristic system in Cartesian coordinates~\eqref{eq:charsysxv} satisfying the initial condition $(X,V)(0,x,v)=(x,v)$.
Differentiating~\eqref{eq:LVP} shows that for any spherically symmetric solution $f=f(t,x,v)$ of the linearised Vlasov-Poisson system,
\begin{equation}\label{eq:num_LVP1}
	\partial_s\left[f(s,(X,V)(s,x,v))\right]=\partial_E\varphi(E(x,v),L(x,v))\,W(s,x,v)\,U_f'(s,R(s,x,v)),\qquad s\in\R,
\end{equation}
where~$R$ and~$W$ are determined by~$X$ and~$V$ via~\eqref{eq:defrwL}; recall the definition of~$L$ and~$E$ in~\eqref{eq:defrwL} and~\eqref{eq:defE}, respectively.
To obtain~\eqref{eq:num_LVP1}, we have used that~$E$ and~$L$ are constant along the characteristic flow~$(X,V)$ and have rewritten the last term on the left-hand side of~\eqref{eq:LVP} using~\eqref{eq:f0varphi} and the spherical symmetries of~$f_0$ and~$U_f$.
Integrating~\eqref{eq:num_LVP1} along the characteristic flow leads to
\begin{equation}\label{eq:num_LVPint}
	f(t,(X,V)(t,x,v))=f(0,x,v)+\partial_E\varphi(E(x,v),L(x,v))\,\int_0^tW(s,x,v)\,U_f'(s,R(s,x,v))\diff s
\end{equation}
for $(t,x,v)\in\R\times\R^3\times\R^3$.
This equation can be interpreted as a new formulation of the linearised Vlasov-Poisson system. 
The existence theory for the system in this form (for isotropic steady states) is studied in~\cite{BaMoRe95}.

To solve~\eqref{eq:num_LVPint} numerically, we approximate this equation by assuming that~$U_f'$ is constant on the time interval $[0,t]$ for $0<t\ll1$, which leads to
\begin{equation}\label{eq:num_LVPintapprox}
	f(t,(X,V)(t,x,v))\approx f(0,x,v)+\partial_E\varphi(E(x,v),L(x,v))\,\int_0^tW(s,x,v)\,U_f'(0,R(s,x,v))\diff s.
\end{equation}
Together with the characteristics~$(X,V)=(X,V)(s,x,v)$, which can be computed numerically as described in the previous section,~\eqref{eq:num_LVPintapprox} allows us to compute $f(t)$ for small values of~$t$. 
Iterating this process yields an approximation of $f(T)$ for arbitrary~$T>0$.

We implement this procedure using a {\em particle-in-cell scheme}, where the key idea is to split the phase space support of the steady state~$f_0$ into finitely many distinct cells. 
We do this by first setting up an equidistant radial grid of step size $\delta r$.
At each fixed radial grid point, the momentum space is segmentated using an equidistant grid in each of the variables~$u\coloneqq|v|$ and~$\alpha\coloneqq\sphericalangle(x,v)=\arccos(\frac{x\cdot v}{|x|\,|v|})$.
Using $(r,u,\alpha)$-coordinates for the initial setup of the cells instead of $(r,w,L)$-coordinates has proven beneficial in previous numerical investigations~\cite{Praktikum20} and the former coordinates are also used in the theoretical study~\cite{Sc87}.
We then place a {\em (numerical) particle} into the $(r,u,\alpha)$-centre of each cell as a representative of the contributions of its cell.
Each particle is assigned its position in $(r,w,L)$-coordinates, the volume of its cell, and the value of the initial distribution $f(0,r,w,L)$.
To compute the corresponding mass density~$\rho_f(0)$, we sum over the contributions of all particles in the momentum variables and interpolate linearly in the radius. 
The mass density is then used to compute the local mass function $m_f(0,r)=4\pi\int_0^rs^2\rho_f(0,s)\diff s$ via Simpson's rule; this numerical scheme takes into account the order of the integrand near the spatial origin $r=0$.
The latter function determines the gravitational force of the perturbation via $U_f'(0,r)=r^{-2}{m_f(0,r)}$. 
By using the boundary condition on~$U_f$ at spatial infinity imposed in~\eqref{eq:poisson}, we can then compute the gravitational potential via $U_f(0,r)=-\int_r^\infty U_f'(0,s)\diff s$.
For all these steps we use the same radial grid as for setting up the cells.
Once these macroscopic quantities are computed, we are in the position to propagate the particles so that they represent the phase space density function $f(\delta t)$ at the next time step $\delta t>0$.
The new positions of the particles are given by evolving the old positions via the characteristic flow~$(X,V)$ of the steady state.
This evolution is achieved as described in the previous section, i.e., by using the fourth-order Runge-Kutta method.
The $f$-values are updated according to~\eqref{eq:num_LVPintapprox}; the integral on the right-hand side containing $U_f'(0)$ is calculated during the computation of the characteristic flow of the steady state.
In fact, we do not use~\eqref{eq:num_LVPintapprox} as it stands (which would correspond to the Euler method) but a slightly refined version of it corresponding to the Runge-Kutta method which is more accurate numerically.
The volumes of the cells associated to the particles remain constant during the particle evolution as the characteristic flow of the steady state is measure preserving~\cite[Lemma~1.2]{Re07}.
Repeating this entire process results in a simulation of the linearised Vlasov-Poisson system.

For the numerical simulations we chose the radial step size $\delta r$ in the order of magnitude $10^{-3}$ to $10^{-4}$ in the case of a steady state with $\Rmax=1$.
The momentum step sizes used to initialise the particles were chosen such that we arrived at a total of $10^7$ to $10^8$ numerical particles; we always made sure to use at least $10^7$ particles.
The proper choices of these parameters depends on the underlying steady state (and the computational resources available); for instance, we used a finer radial grid for steady states with a concentrated region.
The time step size $\delta t$ was chosen to be of a similar magnitude as $\delta r$. 

To be able to perform the simulations within a reasonable computation time, we parallelised the particle propagation as well as various other parts of the algorithm like the computation of the mass density~$\rho_f(t)$.
Fortunately, the particle-in-cell scheme fits very well with parallel computing. 
We use pthreads to implement a shared-memory parallelisation on a CPU based on~\cite{KoRaRe2013}; a GPU-based parallelisation of a related algorithm was developed more recently in~\cite{KoRaWe20}.

In order to evaluate the accuracy of the numerical simulation, we considered several conserved quantities of the linearised Vlasov-Poisson system and monitored the degree to which they remained constant during the simulation.
The first such conserved quantity is the linearised energy~$\Elin$ defined in~\eqref{eq:defElin}.
The second one is the {\em free energy}
\begin{equation}
	\Efree(f(t))\coloneqq-\frac12\int\frac{f(x,v)^2}{\partial_E\varphi(E,L)}\diff(x,v)-\frac1{8\pi}\,\|\partial_xU_{f(t)}\|_{L^2(\R^3)}^2,
\end{equation}
the third one the total mass $M(f(t))$ which is defined similarly to~\eqref{eq:defM0}.
It is straight-forward to verify that the latter quantity is conserved along solutions of the linearised Vlasov-Poisson system.
For the free energy~$\Efree$, this is proven in~\cite[Sc.~4]{BaMoRe95}.
The phase space integrals appearing in these quantities are computed numerically by adding up the contributions of all particles with the respective weights.

During the simulations of the linearised Vlasov-Poisson system for the isotropic polytropes used in Figure~\ref{fig:isopoly_LVP}, the absolute value of the linearised energy was always smaller than $5\cdot10^{-4}$ until the final time $T=50$; notice that $\Elin(f(0))=0$ because $f(0)$ is odd in~$v$.
As already visible in Figure~\ref{fig:isopoly_LVP}, the most difficult case is $k=\frac12$ which is due to the low regularity of the steady state at the boundary of its support.
In the other cases of Figure~\ref{fig:isopoly_LVP}, $|\Elin|$ even remained smaller than $5\cdot10^{-5}$.
The relative error of the free energy, i.e., the absolute value of $\frac{\Efree(f(t))-\Efree(f(0))}{\Efree(f(0))}$, remained smaller than $1\%$ for all simulations in Figure~\ref{fig:isopoly_LVP}. 
The absolute value of the total mass, which also vanishes at $t=0$, stayed smaller than $5\cdot10^{-3}$.
For the other simulations conducted in Section~\ref{ssc:LVPnum}, we ensured that the errors of the conserved quantities are of similar order of magnitudes by appropriately choosing the numerical parameters.
We believe that these errors are sufficiently small to allow us to claim that the numerical simulations indeed accurately describe the behaviour of the solutions of the linearised Vlasov-Poisson system.

\subsection{The pure transport system}\label{ssc:numPuTr}

Simulations of the pure transport system~\eqref{eq:puretransport} can be conducted in a similar but easier way than the simulations of the linearised Vlasov-Poisson system described above.
Concretely, we have to drop the second term on the right-hand side of~\eqref{eq:num_LVPint} to arrive at an alternative form of~\eqref{eq:puretransport}.
As this term caused most of the difficulties in the numerical implementation, it is conceptionally easier to simulate the pure transport equation than the linearised Vlasov-Poisson system.
This manifests itself in the fact that the errors of the conserved quantities of~\eqref{eq:puretransport} -- which are given by the kinetic part of~$\Efree$, the total mass, and~$\Elin$ -- during a numerical simulation of the pure transport equation remain smaller than for the linearised Vlasov-Poisson system (with the same initial configuration and numerical parameters).

\subsection{The Vlasov-Poisson system}\label{ssc:numVP}

Let us now describe how we simulated the (non-linearised) Vlasov-Poisson system~\eqref{eq:vlasov}--\eqref{eq:rho} in spherical symmetry.
Although we focused on initial data~$f(0)$ close to steady states in Section~\ref{ssc:VPnum}, the initial phase space distributions can, in principle, be arbitrary spherically symmetric functions.
As in the two previous sections, we evolve the initial distribution via a {\em particle-in-cell scheme}.
It is proven in~\cite{Sc87} that the simulations presented here indeed converge to the actual solution of the Vlasov-Poisson system if one increases the accuracy of the discretisation.
Similar numerical methods have been used to simulate the radial Vlasov-Poisson system in~\cite{RaRe2018}. 
We have, however, included several improvements which we will discuss below, in particular regarding the treatment of the spatial origin $r=0$ and a higher-order propagation scheme for the particles.
Similar methods are also commonly used to numerically compute solutions of the Vlasov-Poisson system in the plasma physics case~\cite{BiLa91}. 
In particular, in this context, numerous techniques have been developed to further improve the numerical methods used here, see~\cite{Muetal21,MyCoSt17} and the references therein. 
Particle-in-cell methods are also the most common way to simulate the Einstein-Vlasov system in spherical symmetry~\cite{AnRe06,Praktikum20,GueStRe21,Gue23}.

The key idea is again to split the phase space support of the initial distribution~$f(0)$ into finitely many distinct cells.
This step proceeds in the same way as in the linearised case, cf.\ Section~\ref{ssc:numLVP}.
Each of the resulting cells is represented by a {\em (numerical) particle} placed into its centre, which again carries its position in $(r,w,L)$-coordinates, the volume of its cell, and the value of the initial distribution $f(0,r,w,L)$.
These particles represent the initial phase space density~$f(0)$, and the evolution of~$f$ governed by the Vlasov-Poisson system is given through the evolution of the particles.
The particle trajectories are governed by the characteristic system
\begin{equation}\label{eq:numVP_charsys}
	\dot x=v,\qquad \dot v=-\partial_xU_f(t,x),
\end{equation}
where $U_f(t)$ is the gravitational potential generated by~$f$ at time~$t$ via~\eqref{eq:poisson}--\eqref{eq:rho}.
Because~$f$ is only yet known at time $t=0$, we use the following approximation of the above system:
\begin{equation}\label{eq:numVP_charsysapprox}
	\dot x=v,\qquad \dot v=-\partial_xU_f(0,x).
\end{equation}
The right-hand side of this ODE is computed via $\partial_xU_f(0,x)=m_f(0,|x|)\frac x{|x|^3}$, where the local mass function $m_f(0,r)$ is obtained from the particles as described in the context of the linearised system.
Notice that the assumption of spherical symmetry significantly simplifies the computation of the force term in~\eqref{eq:numVP_charsysapprox} because we only have to calculate a one-dimensional integral instead of solving the Poisson equation in three dimensions.
Similar to Section~\ref{ssc:numLVP}, when evolving the particles we deliberately avoid using radial coordinates and instead change to Cartesian $(x,v)$-coordinates to solve the characteristic system.
This technique prevents numerical errors arising from the (artificial) singularity at $r=0$ in radial coordinates and has been developed in~\cite{Praktikum20,RaRe2018}.
Evolving the particles via~\eqref{eq:numVP_charsysapprox} gives a good approximation of their position at the next time step $0<\delta t\ll1$.
In fact, for the particle propagation we do not use~\eqref{eq:numVP_charsysapprox} as it stands (which would correspond to the Euler method), but instead employ a suitable adaptation of the fourth-order Runge-Kutta method similar to~\cite{GueStRe21}.
Also note that the value of~$f$ and the cell size associated to each numerical particle remain constant during the particle evolution because the phase space density~$f$ is constant along the characteristic flow of the Vlasov equation and this flow is measure-preserving~\cite[Lemmas~1.2 and~1.3]{Re07}.
Iterating this entire process results in a simulation of the Vlasov-Poisson system.

We chose numerical parameters similar to the simulations of the linearised Vlasov-Poisson system, cf.\ Section~\ref{ssc:numLVP}.
In particular, we always used between $10^7$ and $10^8$ numerical particles.
To be able to run the program within reasonable time, we employ a shared-memory parallelisation similar to~\cite{Praktikum20,GueStRe21,KoRaRe2013,RaRe2018} based on pthreads.

Similar to the linearised case, we evaluate the accuracies of the simulations by verifying whether conserved quantities of the Vlasov-Poisson system indeed remain constant during the evolution, see~\cite[Sc.~1.5]{Re07} for an overview of such conserved quantities.
The first one is the {\em total energy}
\begin{equation}\label{eq:defEtot}
	\Etot(f(t))\coloneqq\Ekin(f(t))-\frac1{8\pi}\,\|\partial_xU_{f(t)}\|_{L^2(\R^3)}^2,
\end{equation}
where~$\Ekin$ is defined in~\eqref{eq:defEkin}. 
This is the non-linear counterpart to the quantity~$\Elin$ analysed above.
The other conserved quantity we consider here is the total mass $M(f(t))$ which is defined in the same way as before.
These quantities are computed numerically by adding up the contributions of all particles with the respective weights.

For all simulations of the non-linear Vlasov-Poisson system depicted in Figure~\ref{fig:isopoly_VP}, the relative error of the total energy~\eqref{eq:defEtot}, i.e., the absolute value of $\frac{\Etot(f(t))-\Etot(f(0))}{\Etot(f(0))}$, always remained smaller than $5\cdot10^{-4}$.
The relative error of the total mass remained smaller than $8\cdot10^{-5}$.
We ensured that the errors of the simulations of the non-linear Vlasov-Poisson system analysed in Section~\ref{ssc:VPnum} are always of this order of magnitude by appropriately choosing the numerical parameters.
Comparing these errors to the ones encountered when simulating the linearised Vlasov-Poisson system, we conclude that we achieved a higher numerical accuracy at the non-linear level.
Hence, although the linearised Vlasov-Poisson system is easier to analyse from a mathematics point of view, it is not easier to simulate numerically.
This has previously been noted in~\cite[Sc.~3.4]{LeCoBi93}. 
In any case, we are certain that all errors are sufficiently small so that the numerical simulations presented here indeed accurately describe the behaviour of the solutions of the Vlasov-Poisson system.

\makeatletter
\interlinepenalty=10000

\makeatother

\end{document}